\documentclass[pra,floatfix,amsmath,superscriptaddress,twocolumn]{revtex4}

\usepackage{amssymb}
\usepackage{graphicx}
\usepackage{graphics}
\usepackage{amsmath}
\usepackage{amsthm}
\usepackage{color}
\usepackage{bbding}
\usepackage{enumerate}
\usepackage{epstopdf}
\usepackage[mathscr]{eucal}
\usepackage{esvect}
\usepackage[toc,page]{appendix}
\usepackage{subfigure}
\usepackage{float}

\def\bal{\begin{aligned}}
\def\eal{\end{aligned}}
\def\ba{\begin{aligned}}
\def\ea{\end{aligned}}
\def\be{\begin{equation}}
\def\ee{\end{equation}}
\def\beq{\begin{equation}}
\def\eeq{\end{equation}}
\def\bmtl{\begin{multline}}
\def\emtl{\end{multline}}
\def\bestar{\begin{equation*}}
\def\eestar{\end{equation*}}
\def\bea{\begin{eqnarray}}
\def\eea{\end{eqnarray}}
\def\bi{\begin{itemize}}
\def\ei{\end{itemize}}
\def\bc{\begin{center}}
\def\ec{\end{center}}
\def\bt{\begin{tabular}}
\def\et{\end{tabular}}
\def\btm{\begin{theorem}}
\def\etm{\end{theorem}}
\def\barr{\begin{array}}
\def\earr{\end{array}}
\def\bbmat{\begin{bmatrix}}
\def\ebmat{\end{bmatrix}}
\def\bpmat{\begin{pmatrix}}
\def\epmat{\end{pmatrix}}

\def\Order{\mathcal O}
\def\one{\mbox{$1 \hspace{-1.0mm}  {\bf l}$}}

\def\th{\textsuperscript{th} }
\def\gmax{g_{\mathrm{max}}}
\def\epower{\operatorname{e}}

\def\ket#1{\left| #1\right>} 							
\def\bra#1{\left< #1\right|}							
\def\bk#1#2{\left< #1\middle| #2 \right>}				
\def\kb#1#2{\ket{#1}\bra{#2}}  							
\def\bmk#1#2#3{\left< #1\middle| #2 \middle| #3 \right>}  
\def\proj#1{\ket{#1}\bra{#1}}							

\DeclareMathOperator{\tr}{tr}

\newtheorem{theorem}{Theorem}

\def\gmax{g_{\textrm{max}}}

\def\adj#1{#1^{\dagger}}
\def\padj#1{#1^{\phantom{\dagger}}}  

\def\rhoin{\rho_{\mathrm{in}}}

\newcommand{\expect}[1]{\langle #1 \rangle}

\newcommand{\eq}[1]{Eq.~(\ref{eq.#1})}
\newcommand{\sect}[1]{Sec.~\ref{sect.#1}}
\newcommand{\figu}[1]{Fig.~\ref{fig.#1}}

\begin{document}

\author{W. L. Boyajian}
\affiliation{Institute for Theoretical Physics, University of
Innsbruck, Innsbruck, Austria}
\author{B. Kraus}
\affiliation{Institute for Theoretical Physics, University of
Innsbruck, Innsbruck, Austria}

\title{Compressed simulation of thermal and excited states of the 1-D XY-model}

\date{\today}

\begin{abstract}
Since several years the preparation and manipulation of a small number of quantum systems in a controlled and coherent way is feasible in many experiments. In fact, these experiments are nowadays commonly used for quantum simulation and quantum computation. As recently shown, such a system can, however, also be utilized to simulate specific behaviors of exponentially larger systems. That is, certain quantum computations can be performed by an exponentially smaller quantum computer. This compressed quantum computation can be employed to observe for instance the quantum phase transition of the 1D XY--model using very few qubits. We extend here this notion to simulate the behavior of thermal as well as excited states of the 1D XY–-model. In particular, we consider the 1D XY--model of a spin chain of $n$ qubits and derive a quantum circuit processing only $\log(n)$ qubits which simulates the original system. We demonstrate how the behavior of thermal as well as any eigenstate of the system can be efficiently simulated in this compressed fashion and present a quantum circuit on $\log(n)$ qubits to measure the magnetization, the number of kinks, and correlations occurring in the thermal as well as any excited state of the original systems. Moreover we derive compressed circuits to study time evolutions.

\end{abstract}
\maketitle

\section{Introduction}

Simulating a quantum system with a classical computer seems to be an unfeasible task due to the exponential growths of the dimension of the Hilbert space as a function of the number of considered systems. This is why the classical simulation of quantum behavior is usually restricted to a few qubits, although the numerical methods became very powerful. However, as pointed out be Feynman \cite{Fey82} and proven by Lloyd \cite{Llo96} quantum systems can be used to simulate the behavior of the other. The former being such that constituents can be very precisely prepared, manipulated and measured. Many experiments are realizing such a simulation nowadays. Among them experiments utilizing ions in ion--traps, NMR or atoms in optical lattices (see for instance \cite{Blo12,Lanyon11,Houck12} and references therein).

Here we are not concerned about this direct simulation of a quantum system. We are interested in a more economical way of simulating certain quantum behaviors. To this end, we are using the fact that some classes of quantum algorithms, among them those which are based on matchgates, can be simulated classically efficiently. Moreover, it can be shown that matchgate circuits can also be simulated by an exponentially smaller quantum computer \cite{JoKr10}. There, the classical computation is restricted in space such that the computation has to be performed by the quantum computer and cannot be performed by the classical computer. In fact, it has been shown that the computational power of matchgate circuits running on $n$ qubits is equivalent to the one of space-bounded quantum computation with space
restricted to being logarithmic in $n$ \cite{JoKr10}.

As demonstrated in \cite{Kr10, BoMu12} this compressed way of quantum computation can be used to simulated physically interesting behaviours of large systems. To give an example,
consider an experimental set–up, where up to 8 qubits
can be well controlled. Such a set–up can be used to
simulate certain interactions of $2^8= 256$ qubits. In \cite{Kr10, BoMu12}
we demonstrated how the adiabatic evolution of the 1D Ising and more generally the XY-model can be simulated via an exponentially smaller quantum system. More precisely, it is shown there, how the phase
transition of such a model of a spin chain consisting
out of $n$ qubits can be observed via a compressed algorithm processing only $\log(n)$ qubits. The feasibility of such a compressed quantum simulation is due to the fact that the adiabatic evolution and the measurement of the
magnetization employed to observe the phase transition can be described by a matchgate circuit. Remarkably, the number of elementary gates, i.e. the number of single and two–-qubit gates which are required to
implement the compressed simulation can be even smaller than required to implement the original matchgate circuit.
This compressed algorithm has already been experimentally realized using NMR quantum computing \cite{IsingNMR14}. In \cite{BoMu12} we showed that not only the quantum phase transition can be observed in this way, but that various other interesting processes, such as quantum quenching, where the evolution is non–adiabatic, and general time evolutions can
be simulated with an exponentially smaller system.

The aim of this paper is to extend the notion of compressed quantum simulation even further. We will consider the XY-model and derive compressed circuits to simulate the behavior of the thermal and any excited state of the system. To this end, we use the diagonalization of the XY--Hamiltonian presented in \cite{LaVe08}. There, the unitary, which diagonalizes the XY--Hamiltonian has been derived and decomposed into matchgates. Moreover, it has been shown that this unitary can be utilized to generate the thermal and any excited state of the system from a product state.
We use these results to derive compressed circuit which simulate the corresponding matchgate circuits. They can then be used to measure for instance the magnetization of any thermal or excited state of the system using exponentially fewer qubits. Apart from that, we also show that the number of elementary gates will be smaller in the compressed simulation than the direct simulation.

The outline of the remainder of the paper is the following. First, we will recall the definition of matchgates and review how compressed quantum computation can be achieved. Moreover, we will recall some basic properties of the XY-model and its exact diagonalization. In \sect{decomposition} we will recall the results presented in \cite{LaVe08}. There, the unitary, $U$, which transforms the Hamiltonian corresponding to the XY-model to a Hamiltonian of non--interacting systems has been derived. As shown there, this unitary operator can be decomposed into matchgates and can be used to generate the thermal as well as any excited state from a product state. In \sect{compressed} we derive then the compressed circuit corresponding to $U$. We will show that the number of elementary gates required for any of those simulations can be even smaller in the compressed simulation than the original one. Using this circuit it is then easy to derive compressed circuits which can be utilized to simulated the behavior of any eigenstate or thermal state of the model. In \sect{applications} we will then derive compressed circuits for the measurement of correlations in the thermal state, which are related to the number of kinks in the groundstate, as a function of the quenching time and temperature. Moreover, compressed circuits which can be employed to measure the magnetization or correlation of the excited or thermal states will be presented. Finally we will show how time eolution can be simulated with an exponentially smaller system.

\section{Preliminaries}\label{sect.preliminaries}

In this section we first introduce our notation, recall the definition of matchgate circuits, and will then review how these circuits can be compressed. In the last part of this section we recall some basic properties of the XY--model.

\subsection{Notation}

Throughout the paper we denote by $X,Y,Z$ the Pauli operators and by $\one$ the identity operator. The computational basis will be
denoted by $\ket{k}$ for $k\in \{0,1\}^{n}$. We will consider a spin chain of $n$ qubits, where we assume that $n$ is a power of $2$, i.e. $n=2^m$ for some integer $m$. For reasons which will become clear later, we will enumerate the qubits form $0$ to $n-1$. Moreover, we will use the binary notation, $j=[j_k,\dots,j_0]$, for $j=\sum_{l=0}^{n-1} j_l 2^l$.

\subsection{Matchgate circuit and compressed quantum simulation}
Let us recall the concept of matchgate and matchgate circuits. A matchgate is a two--qubit gate which is of the form $ A\oplus B$, where $A$ is acting on span $\{\ket{00},\ket{11}\}$ and $B$ on span $\{\ket{01},\ket{10}\}$. Both, $A$ and $B$ are unitary and have the same determinant. Denoting by
$A_{i,j}$ ($B_{i,j}$) the matrix elements of $A$ ($B$) respectively, we thus have that any matchgate is of the form
\beq
    G(A,B)=
    \bpmat
        A_{1,1}&0&0&A_{1,2}\\
        0&B_{1,1}&B_{1,2}&0\\
        0&B_{2,1}&B_{2,2}&0\\
        A_{2,1}&0&0&A_{2,2}\\
    \epmat,
\eeq
with $\det(A)=\det(B)$. A \emph{matchgate circuit} of size $n$ is defined as a quantum circuit which fulfills the following conditions:
\begin{enumerate}[(i)]
    \item an $n$-qubit quantum state is initially prepared in a computational basis state
    \item it evolves under the action of nearest-neighbor matchgates,
    \item the output is a final measurement on any qubit, say qubit $k$, in the computational basis. That is, the output of the computation is the expectation value $\expect{Z_k}$.
\end{enumerate}

In \cite{valclsim} (see also \cite{jm08,terdiv}) it has been shown that the output of any matchgate circuit can be classically efficiently simulated. We will review now how the classical simulation can be achieved and will then recall the results presented in \cite{JoKr10}, where an equivalence between matchgate circuits and quantum computation running on exponentially less qubit has been shown.

Let us introduce a set of $2n$ Hermitian operators $\{x_j\}_{j=0,\dots,2n-1}$, which satisfy the anticommutation relations
\beq
    \label{eq.Majoranas}
    \{x_j,x_k\}=2\delta_{j,k}\one,\quad\forall j,k.
\eeq
That is, they define a Clifford Algebra, whose elements are any linear combination of products of its $2n$ generators $x_j$. A representation of the generators in term of Pauli operators on a system of $n$ qubits is given by the Jordan-Wigner (JW) representation defined by
\beq
    \label{eq.JordanWigner}
    \bal
        x_{2k}&=Z_0\dots Z_{k-1}X_k\one_{k+1}\dots\one_{n-1},\\
        x_{2k+1}&=Z_0\dots Z_{k-1}Y_k\one_{k+1}\dots\one_{n-1},\\
    \eal
\eeq
for $k=0,\ldots, n-1$. Any matchgate or product of matchgates, $U$, can be written as $U=\epower^{-i\alpha H}$ where $H$ is a quadratic Hamiltonian in the operators $x_j$, that is,
\beq
    \label{eq.quadratichamiltonian}
    H=i\sum_{j\neq k=0}^{2n-1} h_{j,k}x_jx_k.
\eeq
Here, $h$ is a real antisymmetric $2n\times 2n$ matrix \cite{jm08}. In \cite{jm08} it was proven that for any such unitary, $U$, it holds that
\beq
    \label{eq.UxU}
    \adj{U}x_jU=\sum_{k=0}^{2n-1}R_{j,k}x_k,
\eeq
where $R=\epower^{4\alpha h}\in \mathcal{SO}(2n)$ is a real $2n\times 2n$ matrix. Using that $Z_k=-ix_{2k}x_{2k+1}$ [see \eq{JordanWigner}] and \eq{UxU} the outcome of the matchgate circuit can be written as
\beq
    \label{eq.RSR}
    \bal
    \expect{Z_k}&=\bra{0}^{\otimes n} {\adj{U}(-ix_{2k} x_{2k+1})U}\ket{0}^{\otimes n}\\
    &=\sum_{j,l}R_{2k,j}R_{2k+1,l}\bra{0}^{\otimes n}\left(-ix_jx_l\right)\ket{0}^{\otimes n}\\
    &=[RSR^T]_{2k,2k+1},
    \eal
\eeq
where $S_{j,l}=\bra{0}^{\otimes n}\left(-ix_jx_l\right)\ket{0}^{\otimes n}$ for $j\neq l$ and $S_{j,j}=0$. That is $S=\one\otimes iY$, where $\one$ denotes the identity operator on $n=2^{m}$ dimensions. Note that both, $R$ and $S$ are $2n\times 2n$ matrices. Hence, it is evident from Eq. (\ref{eq.RSR}) that $\expect{Z_k}$ can be classically computed efficiently.

Let us now briefly outline how such a computation can be compressed \cite{JoKr10}. Both matrices $R$ and $S$ can be interpreted as operators acting on $m+1$ qubits, which we denote by qubits $0,\dots,m$. Then, $S=iY_{m}$ and the last line in \eq{RSR} can be written as  $\expect{Z_k}=i\bmk{2k}{RY_mR^T}{2k+1}$. Simple algebraic manipulation of this equation leads to
\beq
    \label{eq.compressed}
    \expect{Z_k}=\tr\left[R^T\rhoin(k) R Y_m\right],
\eeq
where $\rhoin(k)=\proj{k}\otimes\proj{+_y}_m$ is a density matrix of $m+1$ qubits. The identity in \eq{compressed} shows that the outcome of the original matchgate circuit, $\expect{Z_k}$, can be measured by preparing a $(m+1)$-qubit system in the state $\rhoin(k)$, evolving it under the action of $R^T$, and then measuring the last qubit in the $Y$ basis. It is important to note here that a decomposition into elementary gates of the compressed gate $R$, which corresponds to the matchgates of the original circuit, can be computed with a classical computer that is bounded to $\Order{(\log{n})}$-space \cite{JoKr10}. This ensures that the classical computer behaves simply as an encoder and that the quantum computer is indeed performing the simulation the original matchgate circuit.

\subsection{XY Hamiltonian}\label{sect.XYHamiltonian}

The 1D--$XY$-Model describes a one-dimensional chain of spins, with nearest-neighbor interactions along the $X$ and $Y$ directions. Moreover, every spin feels the presence of an external magnetic field along the $Z$ direction. The Hamiltonian that governs its dynamics is given by
\beq
    \label{eq.Hamiltonian}
	H=-\left[\sum_{j=0}^{n-1}{g Z_j}+\sum_{j=0}^{n-2}\left(X_jX_{j+1}+\delta Y_j Y_{j+1}\right)\right].
\eeq
The parameter $g$ represents the ratio between the strengths of the external magnetic field and the nearest neighbor interactions, while the factor $\delta\in[0,1]$ represents an anisotropy between the $X$ and $Y$ direction of the spin-spin interactions. The Jordan-Wigner transformation \cite{JordanWigner28} can be used to transform the spin operators into the fermionic operators, $\{c_j\}_{j=0,\dots,n-1}$, with  $c_j=\left(x_{2j}+ix_{2j+1}\right)/2$. Hence, they can be represented in term of Pauli operators as
\beq
    \label{eq.c}
    \bal
        c_j\equiv Z_0\otimes\dots\otimes Z_{j-1}\otimes\sigma^{+}_j\otimes\one_{j+1}\dots \otimes\one_{n-1},
    \eal
\eeq
where $\sigma^{+}_j=\frac{1}{2}\left(X_j+iY_j\right)$. Recall that the fermionic annihilation and creation operators satisfy the anticommutation relations $\{\adj{c}_j,\adj{c}_k\}=\{c_j,c_k\}=0$, $\{c_j,\adj{c}_k\}=\delta_{j,k}$. Clearly, the Hamiltonian $H$ given in Eq. (\ref{eq.Hamiltonian}) is quadratic in the operators $c_i$. It is common to consider the Hamiltonian $H$ with JW boundary conditions, which leads to the Hamiltonian
\beq
    \label{eq.Hbar}
    \bar{H}=H-(X_{n-1}X_{n}+\delta Y_{n-1}Y_{n})
\eeq
where $X_n=\left(Z_0\otimes\dots\otimes Z_{n-1}\right)X_{0}$ and $Y_{n}=\left(Z_0\otimes\dots\otimes Z_{n-1}\right)Y_{0}$. We denote by $H[c]$ the Hamiltonian $\bar{H}$ in terms of the $c$-operators which is given by
\beq
    \label{eq.Hc}
    \bal
        H[c]&=-\sum_{j=0}^{n-1}\Big[g\left(c_j\adj{c_j}-\adj{c_j}c_j\right)\\
            &+(1-\delta)\left(-c_jc_{j+1}+\adj{c_j}\adj{c_{j+1}}\right)\\
            &+(1+\delta)\left(-c_j\adj{c_{j+1}}+\adj{c_j}c_{j+1}\right)\Big].
    \eal
\eeq
Note that the JW boundary conditions translate into periodic boundary conditions in terms of the $c$ operators, that is $c_n=c_0$.

The Hamiltonian $H[c]$ can easily be transformed into a diagonal form \cite{JordanWigner28, LaVe08}. This transformation is divided into two parts: A Fourier transformation and a Bogoliuvov transformation. The Fourier transformation is defined to map the set of operators $\{c_j\}_{j=0,\dots,n-1}$ into the fermionic operators $\{b_j\}_{j=0,\dots,n-1}$ according to
\beq
    \label{eq.Fourier}
    b_k\equiv \frac{1}{\sqrt{n}}\sum_{j=0}^{n-1}{\epower^{i\frac{2\pi}{n}jk}c_j}.
\eeq
Expressing the $c$-operators in \eq{Hc} by the $b$--operators, the Hamiltonian takes the form
\beq
    \label{eq.Hb}
    H[b]=-\sum_{j=0}^{n-1}{\left[2\alpha_j \adj{b_j}b_j+i\beta_{j}\left(b_{-j}b_j+\adj{b_{-j}}\adj{b_j}\right)+g\right]}.
\eeq
Here and in the following we used the definition $b_{-j}\equiv b_{n-j}$, for $j=0,\dots,n-1$. The coefficients $\alpha_j$ and $\beta_j$ are given by
\beq
    \label{eq.alphabeta}
    \bal
        \alpha_j &= \cos\left(\frac{2\pi j}{n}\right)(1+\delta)-g\text{, and}\\
        \beta_j &= \sin\left(\frac{2\pi j}{n}\right)(1-\delta).
    \eal
\eeq

The diagonalization of the Hamiltonian is completed by a Bogoliuvov transformation that maps the operators $\{b_j\}_{j=0,\dots,n-1}$ into a set of fermionic operators $\{a_j\}_{j=0,\dots,n-1}$ which are defined as
\beq
    \label{eq.a}
    a_j\equiv u_jb_j-iv_j\adj{b_{-j}}.
\eeq
Here, the coefficients $u_j \equiv \cos\left( \frac{\theta_j}{2}\right)$ and  $v_j \equiv \sin\left( \frac{\theta_j}{2}\right)$ are functions of the angle
\beq
    \label{eq.theta_j}
    \theta_j=\tan^{-1}\left(\frac{\beta_j}{\alpha_j}\right).
\eeq
Expressing the Hamiltonian in terms of the $a$-operators one finds
\beq
    \label{eq.Ha}
    H[a]=\sum_{j=0}^{n-1}{\epsilon_j \left(\adj{a_j}a_j-\frac{1}{2}\right)}.
\eeq
The energies are determined by
\beq
    \label{eq.epsilons}
    \epsilon_j=-2\left[\alpha_j\cos\left(\theta_j\right)+\beta_j\sin\left(\theta_j\right)\right],
\eeq
e.g. the groundstate energy is $E_0=-\sum_{j=0}^{n-1}\frac{\epsilon_j}{2}$. As \eq{theta_j} defines the angles $\theta_j$ up to the addition of $\pi$, the sign of the energies $\epsilon_j$ is not defined uniquely. In order to have a consistent labeling of the energies for every value of $g$, that is, e.g. that $\epsilon_0$ is the gap between the first exited state and the groundstate energy for any value of $g$, one imposes that the energies $\epsilon_j$ have to be non-negative for all $j$. This constraint determines uniquely the value of $\theta_j$. Note that this choice amounts to keeping the same ordering of the energies, i.e. the first excited state corresponds to the energy $E_0+\epsilon_0$.

\subsubsection{Phase transitions}

For any constant $\delta$ the XY Hamiltonian given in \eq{Hamiltonian} exhibits a quantum phase transition. One distinguished two different scenarios at the phase transition: either the spectrum exhibits a \emph{level crossing} for a finite system size, which occurs if the energy of the groundstate is degenerate at the critical point; or the groundstate energy is not degenerate for any finite system size, but the energy gap between the groundstate and first-excited state tends to zero in the thermodynamic limit. This situation is referred to as an \emph{avoided level crossing}. In both cases the groundstate energy is non-analytic at the phase transition, which is reflected as an abrupt or discontinuous behavior of macroscopic observables, such as the magnetization, at this point. The relevance of quantum phase transitions relies on the fact that they occur at zero temperature and are caused solely by the quantum-mechanical nature of the system under consideration.

The case we consider in this work is the XY Hamiltonian with Jordan-Wigner boundary conditions. Under these boundary conditions, the system exhibits a level crossing at the critical point $g_c$. This is due to the symmetries of the Hamiltonian with respect to parity and momentum \cite{QuantumPhaseTransitions, BoMu12}. The spectrum of any subspace with fixed parity and momentum presents no phase level crossing. However, in any of these subspaces a level crossing is still present when considering the thermodynamic limit.

In \sect{aplicationmagnetization} we present a compressed quantum circuit that can be used to measure the magnetization of the groundstate of the Hamiltonian displaying a phase transition (see also \cite{BoMu12}). This circuit is extended in \sect{applications} where we present a compressed quantum circuit that can be used to measure the magnetization of any thermal and excited state of the Hamiltonian.

\subsubsection{Quantum Quenching}

A relevant physical evolution that can be investigated in systems with quantum phase transitions is denominated quantum quenching. That is, to observe the evolution a the system prepared in the groundstate of the Hamiltonian for a certain value of $g$ when it is driven through the critical point by changing the parameter $g$ abruptly. In the cases where the Hamiltonian does not exhibit a level crossing and the driving speed is slow enough, the system remains in the instantaneous groundstate of the system (adiabatic evolution). However, when the quenching time $t$ is too short, the evolution generates excitations in the system, these excitations can be quantified as a certain amount of defects $\nu$ in the qubit alignment, commonly referred to as \emph{kinks}. The number of kinks, therefore, depends on the quenching speed, and tends to zero when $t\rightarrow \infty$. The form of the dependence of the number of kinks on the quenching speed is predicted by the so called Kibble-Zureck mechanism \cite{Kib03, Zu96}, that predicts a dependence of the form $\nu\propto t^{-1/2}$. The same dependence has been observed for systems of different nature that also present a quantum phase transition (see for instance \cite{ChEr12} and references therein).

In \sect{aplicationquenching} we present a compressed quantum circuit that can be used to measure the number of kinks in the Ising Hamiltonian [$\delta=0$ in \eq{Hamiltonian}] as a function of the quenching speed. There we also compute the dependence of the correlations, which are related to the number of kinks, on the quenching speed for the case of finite temperature.

\section{Matchgate circuit}\label{sect.decomposition}

In the previous section we have seen that the diagonalization of the Hamiltonian is achieved by transforming the set of fermionic operators $\{c_i\}_{i=0,\dots, n-1}$ into a different set of fermionic operators $\{a_i\}_{i=0,\dots, n-1}$. Let us denote by $\mathcal{U}^\dagger$, the corresponding transformation on the fermionic operators. In this section we will review the construction of the corresponding unitary matrix $U$, that transforms the Hamiltonian as
\beq
    \label{eq.UHU}
    UH[a]\adj{U}=H[c].
\eeq

Then, we will recall now the decomposition of $U$ into matchgates presented in \cite{LaVe08} \footnote{Note that this can be easily generalized as for any Hamiltonian $H[c]$ that is a quadratic polynomial in the Majorana fermions [see \eq{quadratichamiltonian}], the unitary $U$ in \eq{UHU} can be decomposed into matchgates \cite{jm08}. Note further that any product of matchgates acting on $n$ qubits can be decomposed into $n^3$ matchgates \cite{jm08}. }.

First, note that due to the JW transformation, one can associate to an $n$-qubit state $\ket{\Psi}$ a fermionic state in the following way,
\beq
    \label{eq.psi}
    \bal
        \ket{\psi\left[a\right]}&=\sum_{i_0,\dots, i_{n-1}=0,1}{\psi[a]_{i_0,\dots,i_{n-1}}\ket{i_0,\dots, i_{n-1}}}\\
                                &=\sum_{i_0,\dots, i_{n-1}=0,1}\psi[a]_{i_0,\dots,i_{n-1}}\\
                                &\phantom{=}\left(\adj{a}_{0}\right)^{i_0}\dots\left(\adj{a}_{n-1}\right)^{i_{n-1}}\ket{\Omega\left[a\right]}.
    \eal
\eeq

Here, and in the following $\ket{\Omega[a]}$ represents the vacuum state of modes $a_i$, that is, $a_i\ket{\Omega[a]}=0$, $\forall i$. Note that using this notation the operator $a_j$ is associated with qubit $j$. As the fermionic operators do not commute, any reordering of the operators might result in a change of signs in Eq. (\ref{eq.psi}). Sometimes, however, it is more convenient to work with a different order, especially as we are seeking for a decomposition of $U$ into nearest neighbor matchgates. We will explain in the following how to take the phases into account.

In order to construct a nearest neighbor matchgate circuit $U$, the transformation $\mathcal{U}$ is decomposed into elementary transformations, that act non-trivially only on 2 fermionic operators, as we will describe in detail in the following subsections. These transformations are conjugated by reordering transformations that permute the position of the qubits. Due to the anticommuting relations of the fermionic operators, swapping, e.g. qubit $j$ and $k\neq j$, yields a phase $(-1)^{i_j i_k+(i_j+i_k)(i_{j+1}+\dots+i_{k-1})}$ in each of the terms of \eq{psi}. It is convenient to  introduce a vector $\boldsymbol{\mu}=\left(\mu_0,\dots \mu_{n-1}\right)$, that is a permutation of the vector $\left(0,\dots,n-1\right)$, whose components indicate which fermionic operator is associated to which qubit. This is done by generalizing the notation of \eq{psi} to
\beq
    \label{eq.psi2}
    \bal
        \ket{\psi\left[y,\boldsymbol{\mu}\right]}&=\sum_{i_0,\dots,i_{n-1}=0,1}\psi[y,\boldsymbol{\mu}]_{i_0,\dots,i_{n-1}}\\
                                            &\phantom{=}\left(\adj{y_{\mu_0}}\right)^{i_0}\dots\left(\adj{y_{\mu_{n-1}}}\right)^{i_{n-1}}\ket{\Omega\left[y\right]}
    \eal
\eeq
where $\left\{y_i\right\}_{i=0,\dots,n-1}$ denotes here an arbitrary set of fermionic operators. Let us denote by $\mathcal{S}_{j,k}$ the fermionic swap operator, which maps $a_j$ to $a_k$ and vice versa. The action of the transformation $\mathcal{S}_{j,k}$ corresponds to the permutation of the two components of $\boldsymbol{\mu}$, $\mu_j$ and $\mu_{k}$. Stated differently, we have $\psi[y,\boldsymbol{\mu}]_{i_0,\dots,i_{n-1}}$ coincides with $\psi[y,\boldsymbol{(0,\ldots,n-1)}]_{i_0,\dots,i_{n-1}}$ up to the sign, which results from the permutation.

In the following section we will consider transformations that act non-trivially only on two fermionic operators, associated to consecutive qubits, e.g. $\{y_{\mu_{2l}},y_{\mu_{2l+1}}\}$. In those cases the corresponding gate acts nontrivially only on qubits $2l$ and $2l+1$
\footnote{
Note that this would not hold if they were not consecutive qubits, but instead arbitrary qubits $j$ and $k>j$. In this case, the corresponding unitary acting on the Hilbert space would act non-trivially on all qubits $j,j+1,\dots,k$.
}.
The correspondence between the computational basis states of qubits $2l$ and $2l+1$ and the fermionic operators is then [see \eq{psi2}],
\beq
    \label{eq.twoqubits}
    \ket{[k_1,k_2][y,\boldsymbol{\mu}]}_{2l,2l+1}=\left(\adj{y}_{\mu_{2l}}\right)^{k_1}\left(\adj{y}_{\mu_{2l+1}}\right)^{k_2}\ket{\Omega[y]}_{2l,2l+1},
\eeq
where, the vacuum state $\ket{\Omega[y]}_{2l,2l+1}$ represents the vacuum state of modes $y_{\mu_{2l}}$ and $y_{\mu_{2l+1}}$.

In the previous section we have followed the conventional procedure used in the literature to diagonalize the XY Hamiltonian. There, the diagonalization has been achieved by applying a Bogoliuvov transformation, $\mathcal{B}$, and a Fourier transformation, $\mathcal{F}$. In the following subsections, we recall how to construct the matchgate circuits corresponding to each of these two transformations \cite{LaVe08}.

\subsection{Bogoliuvov Transformation}\label{sect.bogoliuvovtransformation}

The Bogoliuvov transformation $\mathcal{B}$ maps the fermionic operators $\{a_i\}_{i=0,\dots,n-1}$ into the operators $\{b_i\}_{i=0,\dots,n-1}$, according to \eq{a}. It is easy to see that $\mathcal{B}$ can be written as $\mathcal{B}=\prod_{l=0}^{\frac{n}{2}-1} \mathcal{B}_l$, where  $\mathcal{B}_l$ transforms the operators $\{a_l,a_{-l}\}$ into the operators $\{b_l,b_{-l}\}$ [see \eq{a}], for $l=1,\dots,n/2-1$ and $\mathcal{B}_0$ maps $\{a_0,a_{\frac{n}{2}}\}$ into $\{b_0,b_{\frac{n}{2}}\}$. As the two modes which are mixed by the transformation $\mathcal{B}_l$, have non consecutive indexes we need to reorder the qubits, by applying a permutation $\mathcal{S}_{\mathrm{Bog}}$ to obtain a product of gates acting on consecutive indices. This transformation permutes the position of the qubits in such a way that the qubits $l$ and $-l\equiv n-l$, and $0$ and $\frac{n}{2}$ become consecutive before the transformation $\mathcal{B}_l$ acts, and are returned to the original position afterwards. As was explained above, we can define a vector $\boldsymbol{\mu}$ that indicates the index of each fermionic operator after the reordering transformation [see \eq{psi2}]. We have chosen a particular transformation $\mathcal{S}^{\mathrm{Bog}}$, which corresponds to the mapping of the vector $(0,\ldots,n-1)$ to the vector
\begin{multline}
    \label{eq.lambdabog}
    \boldsymbol{\lambda}^{\mathrm{Bog}}=\Big(0,\frac{n}{2},2,-2,\dots,2l,-2l,\dots,\frac{n}{2}-2,-\frac{n}{2}+2,\\
    \frac{n}{2}-1,-\frac{n}{2}+1,\dots,2l-1,-2l+1,\dots,1,-1\Big).
\end{multline}
Using the following notation for $0\leq l\leq \frac{n}{2}-1$
\beq
    \label{eq.rofj}
    r_l=
    \begin{cases}
     \frac{l}{2}&\text{ for $l$ even,}\\
     \frac{n-1-l}{2}&\text{ for $l$ odd,}
    \end{cases}
\eeq
the components of $\boldsymbol{\lambda}^{\mathrm{Bog}}$ satisfy that $(\lambda^{\mathrm{Bog}}_{2r_l},\lambda^{\mathrm{Bog}}_{2r_l+1})=(l,-l)$ for $1\leq l \leq \frac{n}{2}-1$ and $(\lambda^{\mathrm{Bog}}_0,\lambda^{\mathrm{Bog}}_1)=(0,\frac{n}{2})$. Therefore, it is clear that the transformation $\mathcal{B}_l$ now acts non-trivially on the pair of operators $\Big\{a_{\lambda^{\mathrm{Bog}}_{2r_l}},a_{\lambda^{\mathrm{Bog}}_{2r_l+1}}\Big\}$, which are associated to the consecutive qubits $2r_l$ and $2r_l+1$ [see \eq{psi2}]. Note that both indexes $l$ and $r_l$ are integer numbers in the interval $\left[0,\frac{n}{2}-1\right]$; the index $r_l$ indicates the position of the pair $(l,-l)$ in the vector $\boldsymbol{\lambda}^{\mathrm{Bog}}$.

Let us now compute the gate $B_l$, which corresponds to $\mathcal{B}_l$. First note that the transformation $\mathcal{B}_l$, maps the vacuum state $\ket{\Omega[a]}_{2r_l,2r_l+1}$ into a new vacuum state $\ket{\Omega[b]}_{2r_l,2r_l+1}$. In order to compute this new vacuum state, note that $a_l\ket{\Omega[a]}_{2r_l,2r_l+1}=0$ for $l\in \{\lambda^{\text{Bog}}_{2r_l}, \lambda^{\text{Bog}}_{2r_l+1}\}$. Hence, one can write the projector onto the vacuum state as
\beq
    \proj{\Omega[a]}_{2r_l,2r_l+1}=\padj{a}_{\lambda^{\text{Bog}}_{2r_l}}\adj{a}_{\lambda^{\text{Bog}}_{2r_l}}\padj{a}_{\lambda^{\text{Bog}}_{2r_l+1}}\adj{a}_{\lambda^{\text{Bog}}_{2r_l+1}}.
\eeq
Expressing the operators $a_{\lambda^{\text{Bog}}_{2r_l}}$ and $a_{\lambda^{\text{Bog}}_{2r_l+1}}$ in the $b$ operators [see \eq{a}], the vacuum state $\ket{\Omega[a]}_{2r_l,2r_l+1}$ is mapped to the state
\beq
    \label{eq.omegab}
    \left(u_l\one+iv_l\adj{b}_{\lambda^{\text{Bog}}_{2r_l}}\adj{b}_{\lambda^{\text{Bog}}_{2r_l+1}}\right)\ket{\Omega[b]}_{2r_l,2r_l+1}.
\eeq
It is now straight forward to compute the two-qubit gates $B_l$ that correspond to $\mathcal{B}_l$ by noting that $B_l$ maps $\ket{\left[k_1,k_2\right]\left[a,\boldsymbol{\lambda}^{\mathrm{Bog}}\right]}$ to $\ket{\left[k_1,k_2\right]\left[b,\boldsymbol{\lambda}^{\mathrm{Bog}}\right]}$ for $k_1,k_2\in \{0,1\}$. The relation between the elements of these two bases can be computed using Eqs. (\ref{eq.twoqubits}), (\ref{eq.omegab}) and (\ref{eq.a}), leading to
\begin{multline}
    \label{eq.Bjaction}
    \ket{\left[k_1,k_2\right][a,\boldsymbol{\lambda}^{\mathrm{Bog}}]}_{2r_l,2r_l+1}\\
    \bal
        &=\left(u_l\adj{b}_{\lambda^{\text{Bog}}_{2r_l}}+iv_l b_{\lambda^{\text{Bog}}_{2r_l+1}}\right)^{k_1}\left(u_l\adj{b}_{\lambda^{\text{Bog}}_{2r_l+1}}-iv_l b_{\lambda^{\text{Bog}}_{2r_l}}\right)^{k_2}\\
        &\phantom{=}\left(u_l\one+iv_l\adj{b}_{\lambda^{\text{Bog}}_{2r_l}}\adj{b}_{\lambda^{\text{Bog}}_{2r_l+1}}\right)\ket{\Omega[b]}_{2r_l,2r_l+1}.
    \eal
\end{multline}
By evaluating \eq{Bjaction} for  $k_2,k_1\in \{0,1\}$ and using \eq{twoqubits} on the right-hand side of \eq{Bjaction} to identify the basis states $\ket{[j_1,j_2][b,\boldsymbol{\lambda}^{\mathrm{Bog}}]}_{2r_l,2r_l+1}$, one can compute the components of $\left[B_l\right]_{[j_1,j_2],[k_1,k_2]}=\bk{[j_1,j_2][b,\boldsymbol{\lambda}^{\mathrm{Bog}}]}{[k_1,k_2][a,\boldsymbol{\lambda}^{\mathrm{Bog}}]}$. Doing so, one can easily find that
\beq
    \label{eq.Bj}
    B_l=
    \bpmat
        u_l &0 &0 &iv_l \\
        0 &1 &0 &0 \\
        0 &0 &1 &0 \\
        iv_l &0 &0 &u_l \\
    \epmat.
\eeq
As $u_{l}^2+v_{l}^2=1$, this matrix is a nearest neighbor matchgate that acts non trivially solely on qubits $2r_l$ and $2r_l+1$. Recall that $u_l$ and $v_l$ depends on $\theta_l$, which has to be chosen such that $\epsilon_l\geq 0$ [see \eq{epsilons}]. That is, depending on $g$ the the angle is either $\theta_l$ as given in \eq{theta_j} or $\theta_l+\pi$.

The matrix $B_0$ can be computed similarly, but it requires special attention because as can be seeen in \eq{a}, $a_0=u_0b_0-iv_0\adj{b}_0$ and $a_{\frac{n}{2}}=u_{\frac{n}{2}}b_{\frac{n}{2}}-iv_{\frac{n}{2}}\adj{b}_{\frac{n}{2}}$. That is, the the transformation $\mathcal{B}_0$ mixes the operators with same indexes, which clearly leads to a different gate. In order to compute it let us first note that the coefficients $\alpha_j$ and $\beta_j$ defined in \eq{alphabeta} take the values $\alpha_0=1+\delta-g$ and $\beta_0=0$ for $j=0$. Therefore, the angle $\theta_0=\tan^{-1}\left(\frac{\beta_0}{\alpha_0}\right)$ could take the values $0$ or $\pi$. Using that the energy $\epsilon_0$ is given by $\epsilon_0=-2(1+\delta-g)\cos(\theta_0)$ and that $\theta_0$ is chosen such that $\epsilon_0\geq 0$, we have
\beq
    \theta_0=
    \begin{cases}
        0&\text{ for $g\geq 1+\delta$,}\\
        \pi&\text{ for $g<1+\delta$.}
    \end{cases}
\eeq
It follows from this expression and from \eq{a} that $(a_0,a_{\frac{n}{2}})=(b_0,b_{\frac{n}{2}})$ for $g\geq 1+\delta$ and $(a_0,a_{\frac{n}{2}})=-i(\adj{b}_{\frac{n}{2}},\adj{b}_0)$ for $g<1+\delta$. Thus one can easily compute that
\beq
    \label{eq.B0}
    B_0=
    \begin{cases}
     \one&\text{ for $g\geq 1+\delta$,}\\
        X_0\otimes Z_1&\text{ for $g<1+\delta$.}
    \end{cases}
\eeq
Note that if $g<1+\delta$ the gate $B_0$ is not a matchgate as the condition on the determinants is not satisfied. However, in \sect{compressionB}, when compressing the gates $B_l$, we will show that even in that case a corresponding compressed gate can be found. Note that the appearance of the operator $X_0$ is due to the level crossing of eigenstates with even parity with eigenstates with odd parity. For instance, the groundstate has even parity in the region where $g\geq 1+\delta$ and odd parity in the region where $g<1+\delta$ \footnote{Let us note here that one could also obtain a circuit where all gates are matchgates by defining $B_0=\one\otimes Z_1$ for $g<1+\delta$, and using an initial state where the first qubit is flipped for $g<1+\delta$. As we will see, such a scenario can easily be taken into account when computing the compressed circuit.}.

The circuit $B$ corresponding to the transformation $\mathcal{B}$ is now obtained by multiplying the gates $B_0$ given in \eq{B0} and $B_l$ in \eq{Bj}, that is
\beq
    \label{eq.B}
    B=B_{r_0}\otimes\dots\otimes B_{r_{\frac{n}{2}-1}}.
\eeq

\subsection{Fourier Transformation}\label{sect.fouriertransformation}

The Fourier transformation $\mathcal{F}$ maps the operators $\{b_j\}_{j=0,\dots,n-1}$ into the operators $\{c_j\}_{j=0,\dots,n-1}$, where

\beq
    \label{eq.ck1}
    c_k=\frac{1}{\sqrt{n}}\sum_{j=0}^{n-1}\exp\left(\frac{-i2\pi j k}{n}\right)b_j.
\eeq
Below, we review the procedure presented in \cite{LaVe08} to write the transformation $\mathcal{F}$ as a concatenation of transformations which involve only two fermionic operators at a time. As each of those elementary transformations corresponds to a matchgate, we will obtain in this way the matchgate circuit corresponding to $\mathcal{F}$.

The idea behind the decomposition of $\mathcal{F}$ is that the mapping of the operators $\{b_j\}_{j=0,\dots,n-1}$ into the operators $\{c_j\}_{j=0,\dots,n-1}$ can be accomplished by applying a sequence of maps $\mathcal{F}^{(s)}$, for $0\leq s \leq m-1$, to the operators $b_j$, i.e. $\mathcal{F}=\mathcal{F}^{(m-1)}\circ\dots\circ\mathcal{F}^{(0)}$. The first transformation, $\mathcal{F}^{(0)}$, maps the fermionic operators $\{b_j\}_{j=0,\dots,n-1}$ into a different set of fermionic operators $\{x^{(1)}_j\}_{j=0,\dots,n-1}$, which is defined in such a way that it always mixes only the pair of operators $\{b_j,b_{j'}\}$, where the indices $j$ and $j'$ differ only in the $(m-1)$\th bit of their binary representations. This procedure is repeated iteratively, in such a way that at the $s$\th step, the indices of the pair of operators that are mixed differ only in the $(m-s-1)$\th bit value. Below we explain the procedure in detail.

In the following we will refer to $j=\sum_{l=0}^{m-1}j_l 2^l$ either as the number or its binary representation $[j_{m-1},\dots,j_0]$ whenever it does not lead to any confusion. The phase in \eq{ck1} can now be expanded as
\beq
    \label{eq.phase}
    \epower^{-\frac{i2\pi j k}{n}}=\epower^{-i\pi k\sum_{l=0}^{m-1} 2^{l+1-m}j_l}=\prod_{l=0}^{m-1}\epower^{-\frac{i\pi k j_l}{2^{m-1-l}}}.
\eeq
Let us denote the initial operators $b_j=x^{(0)}_{[j_{m-1},\dots,j_0]}$ ,$\forall j$, to use a notation compatible with the iterative process that we are going to utilize. Using the decomposition of the phases given in \eq{phase}, \eq{ck1} can be written as
\beq
    \label{eq.ck2}
    c_k=\sum_{j_0=0}^{1}\frac{1}{\sqrt{2}}\epower^{-\frac{i\pi k j_0}{2^{m-1}}}\dots\sum_{j_{m-1}=0}^{1}\frac{1}{\sqrt{2}}\epower^{-i\pi k j_{m-1}} x^{(0)}_{[j_{m-1},\dots,j_0]}.
\eeq
Computing only the sum over the index $j_{m-1}$ we obtain
\begin{multline}
    \label{eq.x1}
    \sum_{j_{m-1}=0}^{1}\frac{1}{\sqrt{2}}\epower^{-i\pi k j_{m-1}}x^{(0)}_{[j_{m-1},\dots,j_0]}\\
    \bal
        &=\frac{1}{\sqrt{2}}\left[x^{(0)}_{[0,j_{m-2},\dots,j_0]}+(-1)^{k_0}x^{(0)}_{[1,j_{m-2},\dots,j_0]}\right]\\
        &\equiv x^{(1)}_{[k_0,j_{m-2},\dots,j_0]},
    \eal
\end{multline}
where $k_0$ is the last bit value of $k$. Recalling that $[k_0,j_{m-1},\dots,j_0]$ corresponds to the number $k_02^{m-1}+\sum_{l=0}^{m-2}j_l2^{l}$, one can see that \eq{x1} defines a set of $n$ fermionic operators $\{x^{(1)}_i\}_{i=0,\dots,n-1}$ as a function of the operators $\{x^{(0)}_i\}_{i=0,\dots,n-1}$. Explicitly, we have that
\beq
    \label{eq.x1b}
    \bal
        x^{(1)}_{[0,j_{m-1},\dots,j_0]}&=\frac{1}{\sqrt{2}}\left[x^{(0)}_{[0,j_{m-2},\dots,j_0]}+x^{(0)}_{[1,j_{m-2},\dots,j_0]}\right]\\
        x^{(1)}_{[1,j_{m-1},\dots,j_0]}&=\frac{1}{\sqrt{2}}\left[x^{(0)}_{[0,j_{m-2},\dots,j_0]}-x^{(0)}_{[1,j_{m-2},\dots,j_0]}\right].\\
    \eal
\eeq
Let us call the transformation that maps the original fermionic operators $\{x^{(0)}_i\}_{i=0,\dots,n-1}$ into the fermionic operators $\{x^{(1)}_i\}_{i=0,\dots,n-1}$, according to \eq{x1b}, $\mathcal{F}^{(0)}$. Note that the transformation $\mathcal{F}^{(0)}$ mixes pairs of operators whose indices differ only in the $(m-1)$\th bit value. Replacing \eq{x1} in \eq{ck2} we obtain
\beq
    \label{eq.ck3}
    \bal
        c_k &=\sum_{j_0=0}^{1}\frac{1}{\sqrt{2}}\epower^{-\frac{i\pi k j_0}{2^{m-1}}}\dots\\
        &\phantom{=}\sum_{j_{m-2}=0}^{1}\frac{1}{\sqrt{2}}\epower^{-\frac{i\pi k j_{m-2}}{2}}x^{(1)}_{[k_0,j_{m-2},\dots,j_0]}.
    \eal
\eeq

Repeating this procedure $s$ times, for $s \in \{0,\dots,m-1\}$, we find
\begin{multline}
    \label{eq.ck4}
    c_k=\sum_{j_0=0}^{1}\frac{1}{\sqrt{2}}\epower^{-\frac{i\pi k j_0}{2^{m-1}}}\dots\\
    \sum_{j_{m-s-1}=0}^{1}\frac{1}{\sqrt{2}}\epower^{-\frac{i\pi k j_{m-s-1}}{2^s}}x^{(s)}_{[k_0,\dots,k_{s-1},j_{m-s-1},\dots,j_0]}.
\end{multline}
Similarly as before, we define the operators $\{x^{(s+1)}_i\}_{i=0,\dots,n-1}$ as the sum over the index $j_{m-s-1}$ in \eq{ck4}, that is
\begin{multline}
    \label{eq.xs1}
    x^{(s+1)}_{[k_0,\dots,k_s,j_{m-s-2},\dots,j_0]}\equiv\frac{1}{\sqrt{2}}\Big(x^{(s)}_{[k_0,\dots,k_{s-1},0,j_{m-s-2},\dots,j_0]}+\\
    (-1)^{k_s}\epower^{-i\pi\tilde{\varphi}(k,s)}x^{(s)}_{[k_0,\dots,k_{s-1},1,j_{m-s-2},\dots,j_0]}\Big).
\end{multline}
In this general case, the phase difference between the operators $x^{(s)}_{[k_0,\dots,k_{s-1},0,j_{m-s-2},\dots,j_0]}$ and $x^{(s)}_{[k_0,\dots,k_{s-1},1,j_{m-s-2},\dots,j_0]}$ has an additional component $\tilde{\varphi}(k,s)$. It can be extracted by evaluating explicitly $\epower^{-\frac{i\pi k j_{m-s-1}}{2^s}}$ for $j_{m-s-1}=1$, that is
\beq
    \label{eq.phikl}
    \epower^{-\frac{i \pi k }{2^s}}=\epower^{-i\pi\sum_{l=0}^{m-1}2^{l-s}k_l}=(-1)^{k_s}\epower^{-i\pi\tilde{\varphi}(k,s)},
\eeq
with $\tilde{\varphi}(k,s)=\sum_{l=0}^{s-1}2^{l-s}k_l$. In the last equality we used the fact that the terms in the sum with $l>s$ lead to trivial phases. In the same way as $\mathcal{F}^{(0)}$ was defined via \eq{x1b}, we define the transformation $\mathcal{F}^{(s)}$ as the one that maps the fermionic operators $\{x^{(s)}_i\}_{i=0,\dots,n-1}$ into the new fermionic operators $\{x^{(s+1)}_i\}_{i=0,\dots,n-1}$ given in \eq{xs1}. The transformation mixes pairs of operators whose indices differ only in the $(m-s-1)$\th bit value.

The last transformation of the succession $\mathcal{F}=\mathcal{F}^{(m-1)}\circ\dots\circ\mathcal{F}^{(0)}$, i.e. $\mathcal{F}^{(m-1)}$, maps the fermionic operators $\{x^{(m-1)}_i\}_{i=0,\dots,n-1}$ into the fermionic operators $\{x^{(m)}_i\}_{i=0,\dots,n-1}$. The output of this transformation coincides, up to a permutation of the indices, with the desired set of operators $\{c_i\}_{i=0,\dots,n-1}$. In fact, setting $s=m-1$ in \eq{ck4} we obtain
\beq
    \label{eq.ckequivalence}
    c_{[k_{m-1},\dots,k_0]}=x^{(m)}_{[k_0,\dots, k_{m-1}]}.
\eeq

Let us now have a closer look at the transformation $\mathcal{F}^{(s)}$ for an arbitrary $s$. Due to \eq{xs1} we have that the transformation $\mathcal{F}^{(s)}$ can be written as $\mathcal{F}^{(s)}=\mathcal{F}^{(s)}_0\circ\dots\circ\mathcal{F}^{(s)}_{\frac{n}{2}-1}$, where each map $\mathcal{F}^{(s)}_l$, for $0\leq l \leq \frac{n}{2}-1$, acts non-trivially only on one pair of operators, namely, $\left\{x_{r_{l,s}}^{(s)},x_{r_{l,s}+2^{m-s-1}}^{(s)}\right\}$, where $r_{l,s}$ corresponds to the number whose binary representation is
\beq
    \label{eq.rl}
    \left[l_{m-2},\dots,l_{m-s-1},0,l_{m-s-2},\dots,l_0\right].\\
\eeq
Hence, for each $l$, the indices of the two operators that are non-trivially transformed, differ by $2^{m-s-1}$, i.e. by their $(m-s-1)$\th bit value in their binary representation. More explicitly we have that $\mathcal{F}^{(s)}_l$ maps the pairs of operators $\left\{x^{(s)}_{r_{l,s}},x^{(s)}_{r_{l,s}+2^{m-s-1}}\right\}$, into the pairs of operators $\left\{x^{(s+1)}_{r_{l,s}},x^{(s+1)}_{r_{l,s}+2^{m-s-1}}\right\}$ according to
\beq
    \bal
        \label{eq.xs1b}
        x^{(s+1)}_{r_{l,s}}&=\frac{1}{\sqrt{2}}\left(x^{(s)}_{r_{l,s}}+\epower^{-i\pi\varphi(l,s)}x^{(s)}_{r_{l,s}+2^{m-s-1}}\right)\\
        x^{(s+1)}_{r_{l,s}+2^{m-s-1}}&=\frac{1}{\sqrt{2}}\left(x_{r_{l,s}}^{(s)}-\epower^{-i\pi\varphi(l,s)}x^{(s)}_{r_{l,s}+2^{m-s-1}}\right),\\
    \eal
\eeq
where the phase is given by
\beq
    \label{eq.varphi}
    \varphi(l,s)=\sum_{j=0}^{s-1}2^{j-s}l_{m-2-j}.
\eeq
Note that this is just a redefinition of the phase $\tilde{\varphi}(k,s)$, given in \eq{phikl}, that takes into account that the binary bits of $l$ in \eq{rl} are sorted differently than those of $k$ in \eq{ck4}.

Now we are in the position to review how to construct the matchgate associated to the elementary transformations $\mathcal{F}^{(s)}_l$, $\forall l,s$. Let us assume that before and after each of these operations, appropriate reordering operations are implemented such that the ordering of the fermionic operators is described by a vector $\boldsymbol{\lambda}^{(s)}$ [see \eq{psi}], with components $\lambda^{(s)}_{2l}=r_{l,s}$ and $\lambda^{(s)}_{2l+1}=r_{l,s}+2^{m-s-1}$, for $l=0,\dots,\frac{n}{2}-1$. Therefore the transformation $\mathcal{F}^{(s)}_l$ maps now the operators $\left\{x^{(s)}_{\lambda^{(s)}_{2l}},x^{(s)}_{\lambda^{(s)}_{2l+1}}\right\}$ into $\left\{x^{(s+1)}_{\lambda^{(s)}_{2l}},x^{(s+1)}_{\lambda^{(s)}_{2l+1}}\right\}$, that are associated to consecutive qubits $2l$ and $2l+1$. Hence, the gate corresponding to $\mathcal{F}^{(s)}_l$, which we denote by $F_l^{(s)}$, acts non trivially only on the $2l$\th and $(2l+1)$\th qubits. Unlike the Bogoliuvov transformation, this transformation is passive and therefore does not modify the vacuum state, i.e. $\ket{\Omega[x^{(s+1)}]}=\ket{\Omega[x^{(s)}]}$
\footnote{In order to see that, we proceed as in the previous section and write the projection onto the vacuum state $ \ket{\Omega[x^{(s+1)}]}_{2l,2l+1}$ as $\proj{\Omega[x^{(s+1)}]}_{2l,2l+1}=x^{(s+1)}_{\lambda^{(s)}_{2l}}x^{(s+1)\dagger}_{\lambda^{(s)}_{2l}}x^{(s+1)}_{\lambda^{(s)}_{2l+1}}x^{(s+1)\dagger}_{\lambda^{(s)}_{2l+1}}.$
Using \eq{xs1b} to write the transformed operators $\left\{x^{(s+1)}_{\lambda^{(s)}_{2l}},x^{(s+1)}_{\lambda^{(s)}_{2l+1}}\right\}$ in terms of the operators $\left\{x^{(s)}_{\lambda^{(s)}_{2l}},x^{(s)}_{\lambda^{(s)}_{2l+1}}\right\}$, we find that
$\ket{\Omega[x^{(s+1)}]}_{2l,2l+1}=\ket{\Omega[x^{(s)}]}_{2l,2l+1}.$
}.
The two-qubit gate $F_l$ that corresponds to the transformation $\mathcal{F}_l$ can be computed by noting that $F_l$ maps $\ket{[k_1,k_2][x^{(s)},\boldsymbol{\lambda}^{(s)}]}_{2l,2l+1}$ to $\ket{[j_1,j_2][x^{(s+1)},\boldsymbol{\lambda}^{(s)}]}_{2l,2l+1}$. Using the notation given in \eq{twoqubits} with operators $x^{(s)}$, and \eq{xs1b} one obtains that
\begin{multline}
    \ket{[k_1,k_2][x^{(s)},\boldsymbol{\lambda}^{(s)}]}_{2l,2l+1}=\left(\frac{x_{\lambda^{(s)}_{2l}}^{(s+1)\dagger}+x_{\lambda^{(s)}_{2l+1}}^{(s+1)\dagger}}{\sqrt{2}}\right)^{k_1}\\
    \left[\epower^{i\pi\varphi(l,s)}\frac{\left(x_{\lambda^{(s)}_{2l}}^{(s+1)\dagger}-x_{\lambda^{(s)}_{2l+1}}^{(s+1)\dagger}\right)}{\sqrt{2}}\right]^{k_2}\ket{\Omega[x^{(s+1)}]}_{2l,2l+1}.
\end{multline}
By evaluating this expression for $k_1,k_2\in \{0,1\}$, and using \eq{twoqubits} to identify the basis elements $\ket{[j_1,j_2][x^{(s+1)},\boldsymbol{\lambda}^{(s)}]}_{2l,2l+1}$ on the right-hand side of the equation, one can compute the matrix elements $\left[F_l\right]_{[j_1,j_2],[k_1,k_2]}=\bk{[j_1,j_2][x^{(s+1)},\boldsymbol{\lambda}^{(s)}]}{[k_1,k_2][x^{(s)},\boldsymbol{\lambda}^{(s)}]}$. One can easily verify that, $F_l^{(s)}$, which is a matchgate acting non trivially only on qubits $2l$ and $2l+1$, is given by
\beq
    \label{eq.Fjs}
    F_l^{(s)}=
    \bpmat
        1&0&0&0\\
        0&-\frac{\alpha(l,s)}{\sqrt{2}}&\frac{1}{\sqrt{2}}&0\\
        0&\frac{\alpha(l,s)}{\sqrt{2}}&\frac{1}{\sqrt{2}}&0\\
        0&0&0&-\alpha(l,s)\\
    \epmat,
\eeq
where $\alpha(l,s)=\epower^{i\pi\varphi(l,s)}$ with $\varphi(l,s)$ given in \eq{varphi}. As $\mathcal{F}^{(s)}$ is decomposed as a sequence of transformations $\mathcal{F}^{(s)}_l$ for $0\leq l \leq \frac{n}{2}-1$, the matchgate product associated to $\mathcal{F}^{(s)}$ is then given by
\beq
    \label{eq.Fs}
    F^{(s)}=F_0^{(s)}\otimes\dots\otimes F_{\frac{n}{2}-1}^{(s)}.
\eeq
The matchgate circuit corresponding to the whole Fourier transformation is then given by the product of circuits $F^{(s)}$, conjugated by appropriate reordering matchgates that correspond to the reordering transformations discussed above.

\subsection{Reordering operations}\label{sect.reordering}

As mentioned before, the different steps of the transformation which diagonalizes the Hamiltonian have to be conjugated by reordering transformations. The reason for that is that in order to obtain nearest-neighbor matchgate circuits corresponding to the Bogoliuvov and Fourier transformations, only fermionic operators in consecutive positions have to be mixed at any step of the transformations. The transformation $\mathcal{U}$ is thus given by
\footnote{
    Note that two reshuffling operations appear together, that is $\mathcal{S}^{(m+1)}$ and $\mathcal{S}^{(m)}$, and could be combined into a new operation $\mathcal{S}'$. However we leave it in this way, since $\mathcal{S}^{(m)}$ has a simple interpretation, and $\mathcal{S}^{(m+1)}$ has to be computed anyways since it appears in the first position again.
}
\beq
    \label{eq.mathcalU}
    \mathcal{U}=\mathcal{S}^{(m)}\mathcal{F}^{(m-1)}\mathcal{S}^{(m-1)}\circ\dots\circ \mathcal{F}^{(0)}\mathcal{S}^{(0)}\left(\mathcal{S}^{\mathrm{Bog}}\right)^{-1}\mathcal{B}\mathcal{S}^{\mathrm{Bog}},
\eeq
where the transformations $\mathcal{S}^{\mathrm{Bog}}$ and $\mathcal{S}^{(k)}$, for $0\leq k\leq m$ represent the reordering transformations. Obviously, any reordering transformation can be accomplished by a sequence of reordering transformations $\mathcal{S}_{l,l+1}$ that correspond to the swapping of qubits $l$ and $l+1$ respectively. In Appendix \ref{ap.appendixA} we derive the explicit form for each of the permutations. Here, we just recall that the gate, $S_{l,l+1}$, corresponding to $\mathcal{S}_{l,l+1}$, which can be computed as before, corresponds to the fermionic swap, i.e.
\beq
    \label{eq.Sjj1}
    S_{l,l+1}=
    \bpmat
        1 &0 &0 &0 \\
        0 &0 &1 &0 \\
        0 &1 &0 &0 \\
        0 &0 &0 &-1 \\
    \epmat.
\eeq

The gate $S_{j,k}$, corresponding to $\mathcal{S}_{j,k}$ which exchanges $\left(a_k,\adj{a}_k\right)$ with $\left(a_j,\adj{a}_j\right)$ is then given by
\beq
    \label{eq.Sjk}
    S_{j,k}=\prod_{l=j}^{k-1} S_{l,l+1}.
\eeq
The matchgate circuit $U$ can now be written as
\beq
    \label{eq.Ucircuit}
    U=S^{(m)}F^{(m-1)}S^{(m-1)}\cdots F^{(0)}S^{(0)}\adj{S}_{\mathrm{Bog}}B\padj{S}_{\mathrm{Bog}},
\eeq
where $F^{(s)}$ is defined in \eq{Fs}, $B$ in \eq{B} and the gates $S$ are given in the Appendix \ref{ap.appendixA} as a function of the fermionic swaps $S_{j,k}$. In the next section, we derive the compressed gate corresponding to $U$ using \eq{UxU}.

\section{Compressed circuit}\label{sect.compressed}

As shown in \sect{decomposition}, the matchgate circuit $U$ is decomposed into matchgates $B_l$, $F_l^{(s)}$ and $S_{j,k}$ for $l \in \{0,\ldots, \frac{n}{2}-1\}$ and $j,k \in \{0,\ldots,n-1\}$. In this section we construct the compressed gates corresponding to each of these gates, namely $R_{B_l}$, $R_{F_l}$ and $R_{S_{j,k}}$. This is done by following the procedure described in \sect{preliminaries}. That is, we write an arbitrary matchgate $G$ in the form $G=\epower^{-iH_G}$, with $H_G$ of the form of \eq{quadratichamiltonian}, from where we can extract the matrix of coefficients $h_G$. The corresponding compressed gate is then given by $R_G=\epower^{4 h_G}$.

Although $R_{B_l}$, $R_{F^{(s)}_l}$ and $R_{S_{j,k}}$ are all the compressed gates required to construct the compressed circuit $R$, it is possible to use much less gates by taking the symmetries of the Hamiltonian into account. As we will see, the compressed gate corresponding to the whole product of matchgates $F^{(s)}=\prod_{l=0}^{\frac{n}{2}-1}F^{(s)}_l$ given in \eq{Fs} can be more economically implemented than the product of the compressed gates corresponding to $F_l^{(s)}$, $\forall l$. The same applies to any of the factors of the product given in \eq{Ucircuit}. For this reason, we write down the compressed gate corresponding to $U$, as the product
\beq
    \label{eq.Rcircuit}
    R=R_{S^{(m)}}R_{F^{(m-1)}}R_{S^{(m-1)}}\dots R_{F^{(0)}}R_{S^{(0)}}R_{S^{\mathrm{Bog}}}^{T}R_BR_{S^{\mathrm{Bog}}},
\eeq
where we have used the notation $R_V$ to denote the compressed gate corresponding to a unitary $V$. We omitted in \eq{Rcircuit} the dependence of $R$ on $g$ and $\delta$. However, whenever we want to stress this dependence, we will write $R(g,\delta)$ for $R$ given in \eq{Rcircuit}. Below we are going to provide the expression for each of the factors.

\subsection{Compression of the Bogoliuvov transformation}\label{sect.compressionB}

In this subsection we derive the compressed gate, $R_{B}$, corresponding to the Bogoliuvov transformation $B$, which is a product of the matchgates $B_l$ [see \eq{B}]. In order to do so we compute $R_{B_{l}}$, i.e. the compressed gate corresponding to $B_l$.

We consider first the case $g\geq 1+\delta$. In this regime $B_0=\one$ [see \eq{B0}] and therefore we can extend the definition of $B_l$ in \eq{Bj} to the case $l=0$. As explained in \sect{bogoliuvovtransformation}, the gate $B_l$ acts on qubits $2r_l$ and $2r_l+1$, where $r_l$ is given in \eq{rofj}. It is easy to see that the gate $B_l$ can be written as
\beq
    \label{eq.Bj2}
    B_l=\epower^{i\frac{\theta_l}{4}(X_{2r_l}X_{2r_l+1}-Y_{2r_l}Y_{2r_l+1})},
\eeq
where $\theta_l$ is given in \eq{theta_j}. Using the Jordan-Wigner representation given in \eq{JordanWigner}, \eq{Bj2} becomes $B_l=\epower^{iH_{B_l}}$, with a quadratic Hamiltonian
\beq
    H_{B_l}=-i\frac{\theta_l}{8}\left(x_{4r_l}x_{4r_l+3}+x_{4r_l+1}x_{4r_l+2}-\text{h.c.}\right),
\eeq
where $\text{h.c.}$ denotes here and in the following the hermitian conjugation of the preceding terms. Writing $H_{B_l}=i\sum_{j,k}\left[h_{B_l}\right]_{j,k}x_{j}x_{k}$, with $h_{B_l}=-\frac{\theta_l}{8}\left(\kb{4r_l}{4r_l+3}+\kb{4r_l+1}{4r_l+2}-\text{h.c.}\right)$ one can easily compute the compressed gate $R_{B_l}=\epower^{4h_{B_l}}$, which is then given by
\beq
    R_{B_l}=\one\oplus
    \bpmat
        \cos\left(\frac{\theta_l}{2}\right) & 0&0& -\sin\left(\frac{\theta_l}{2}\right)\\
        0&\cos\left(\frac{\theta_l}{2}\right)&-\sin\left(\frac{\theta_l}{2}\right)&0\\
        0&\sin\left(\frac{\theta_l}{2}\right)&\cos\left(\frac{\theta_l}{2}\right)&0\\
        \sin\left(\frac{\theta_l}{2}\right) & 0&0& \cos\left(\frac{\theta_l}{2}\right)\\
    \epmat.
\eeq
Here, $R_{B_l}$ acts non--trivially only on the basis elements $\{\ket{4r_l},\dots,\ket{4r_l+3}\}$. Note that $R_{B_l}=\adj{V_{B_l}}D_{B_l}V_{B_l}$, where $D_{B_l}$ is diagonal and the structure of $V_{B_l}$ is independent of $l$ (see below). Due to the independency of $V_{B_l}$ on $l$ (only the basis elements on which it acts non--trivially depend on $l$), $R_B$ can be written as
\beq
    \label{eq.RB}
    R_{B}=\prod_{l=0}^{\frac{n}{2}-1}R_{B_l} =\adj{V}_{B} D_{B} V_{B}.
\eeq
Here, the gate $D_B$ is diagonal and is given by
\begin{multline}
    \label{eq.DB}
    D_B=\prod_{l=0}^{\frac{n}{2}-1}D_{B_l}\\
    =\sum_{l=0}^{\frac{n}{2}-1}\Big[\proj{r_l}\otimes\big(\epower^{-i\theta_l}\proj{0}+\epower^{i\theta_l}\proj{1}\big)\Big]\otimes\one_m,
\end{multline}
and the gate
\beq
    V_B=\prod_{l=0}^{\frac{n}{2}-1}V_{B_l}=\one\otimes\frac{1}{\sqrt{2}}
    \bpmat
        i&0&0&1\\
        0&i&1&0\\
        0&-i&1&0\\
        -i&0&0&1\\
    \epmat,
\eeq
acts non-trivially only on the last two qubits and it is independent of the angles $\theta_l$.

In the regime where $g<1+\delta$ $B_0$ is given by $X_0\otimes Z_1$ [see \eq{B0}]. Even though it cannot be written as the exponent of a quadratic Hamiltonian one can nevertheless find an expression similar to \eq{UxU} as we will show in the following. Note that $B_0$ commutes with $x_j$ for $j=0,2,3$ and anticommutes with any other $x_j$. Thus we have that
\beq
    \label{eq.B0xB0}
    \adj{B}_0x_j B_0=\sum_{k=0}^{2n-1} [R_{B_0}]_{j,k}x_k,
\eeq
with
\beq
    \label{eq.RB0}
    R_{B_0}=2\big(\kb{0}{0}+\kb{2}{2}+\kb{3}{3}\big)-\one.
\eeq
In contrast to the matrix $R$ in \eq{UxU}, the matrix $R_{B_0}$ is orthogonal but its determinant is not one. However, \eq{B0xB0} ensures classical simulability and compressibility of the matchgate circuit, independently of the determinant of $R_{B_0}$ \footnote{Note that the decomposition of $R_{B_0}$ into elementary gates works in the same way as has been done in \cite{JoKr10} for matchgates. Another method would be to multiply $R_{B_0}$ with a phase so that it would be in SO(2n).}.

In summary, for $g\geq 1+\delta$, the compressed gate corresponding to the Bogoliuvov transformation is given by $R_{B}$ given in \eq{RB}, while for $g<1+\delta$, it is given by the product $R_{B}R_{B_0}$.

\subsection{Compression of the Fourier transformation}

In this section we compute the compressed gate $R_{F_l^{(s)}}$ corresponding to the matchgate $F_l^{(s)}$. The compressed circuit corresponding to $F^{(s)}$ is then given by the product $R_{F^{(s)}}=\prod_{l=0}^{\frac{n}{2}-1} R_{F^{(s)}_l}$. Below we derive $R_{F^{(s)}}$ which is computed similarly to $R_B$ in \eq{RB}.

The gate $F^{(s)}_l$ given in \eq{Fjs} acts non--trivially on qubits $2l$ and $2l+1$. The corresponding compressed gate can be easily found to be
\beq
    \label{eq.RFjs}
    R_{F_l^{(s)}}=R_{G_l}\adj{V}_{F^{(s)}_l}D_{F^{(s)}_l}V_{F^{(s)}_l},
\eeq
where $D_{F_l^{(s)}}$ is diagonal and the structure of the matrices $R_{G_l}$ and $V_{F^{(s)}_l}$ does neither depend on $l$ nor on $s$. The matrix $R_{F_l^{(s)}}$, and therefore all the elements in the decomposition given in \eq{RFjs} act non trivially only on the basis elements $\left\{\ket{4l},\dots,\ket{4l+3}\right\}$ [see \eq{UxU}]. Combining all that leads to
\beq
    \label{eq.RFsdecomposition}
    R_{F^{(s)}}=\prod_{l=0}^{\frac{n}{2}-1} R_{F^{(s)}_l}=R_{G} V_{F}D_{F^{(s)}}\adj{V}_{F}.
\eeq
Here, the gate
\begin{multline}
    \label{eq.DFs}
    D_{F^{(s)}}=\proj{0}_{m-1}\otimes\one+\proj{1}_{m-1}\otimes\\
    \sum_{l=0}^{\frac{n}{2}-1}\left[\proj{l}\otimes\big(\epower^{-i\pi\varphi(l,s)}\proj{0}_m+\epower^{i\pi\varphi(l,s)}\proj{1}\big)\right],
\end{multline}
is diagonal and the gates
\beq
    \bal
         R_G&=\prod_{l=0}^{\frac{n}{2}-1} R_{G_l}=\one\otimes\frac{1}{\sqrt{2}}
        \bpmat
                1&0&1&0\\
                0&1&0&1\\
                -1&0&1&0\\
                0&-1&0&1\\
        \epmat
        \text{ and }\\
        V_{F}&=\prod_{l=0}^{\frac{n}{2}-1} V_{F^{(s)}_l}=\one\otimes
        \bpmat
            1&0&0&0 \\
            0&1&0&0\\
            0&0&\frac{-i}{\sqrt{2}}&\frac{1}{\sqrt{2}}\\
            0&0&\frac{i}{\sqrt{2}}&\frac{1}{\sqrt{2}}\\
        \epmat,
    \eal
\eeq
act non-trivially only on the last two qubits.

\subsection{Compression of the reordering steps}

As mentioned in the previous section, every reordering step consists of a product of permutations that are associated to a product of the matchgates $S_{j,k}$ given in \eq{Sjk}. Below we provide its corresponding compressed gate $R_{S_{j,k}}$. The compressed gates in \eq{Rcircuit} that correspond to each reordering step can all be computed as a product of matrices $R_{S_{j,k}}$ (for the detailed circuits see Appendix \ref{ap.appendixB}).

The matrix $S_{j,j+1}$ [see \eq{Sjj1}] can be written, up to a global phase, as
\beq
    {S}_{j,j+1}=\epower^{\frac{\pi}{8}\left(x_{2j}x_{2j+1}+x_{2j+2}x_{2j+3}+x_{2j+1}x_{2j+2}-x_{2j}x_{2j+3}-\text{h.c.}\right)},
\eeq
where the operators $x_j$ are given in \eq{JordanWigner}. Hence, the corresponding compressed gate is given by
\beq
    \label{eq.RSjj1}
    \bal
        R_{{S}_{j,j+1}}&=\Big(\one+\kb{j}{j+1}+\kb{j+1}{j}\\
        &\phantom{=}-\proj{j}-\proj{j+1}\Big)\otimes\one_{m},
    \eal
\eeq
where the operator in parenthesis is acting on the qubits $0,\dots,m-1$. As $S_{j,k}$ can be easily obtained by multiplying consecutive fermionic swap gates, $S_{j,j+1}$, the compressed gate, $R_{S_{j,k}}$, can be obtained by multiplying consecutive gates of the form of \eq{RSjj1}. Explicitly we have
\beq
    \label{eq.RSjk}
    R_{{S}_{j,k}}=\big(\one+\kb{j}{k}+\kb{k}{j}-\proj{j}-\proj{k}\big)\otimes\one_{m}.
\eeq
In the subsequent section we determine the number of elementary gates required to implement the compressed circuit.

\subsection{Comparison of the number of elementary gates}

As mentioned at the beginning of this section, the symmetries of the Hamiltonian allow to decrease the number of single- and two-qubit gates required to implement the compressed gate $R$ compared to the number of nearest neighbor matchgates required to implement $U$ in the original matchgate circuit. While the decomposition of $U$ into matchgates considered here requires $\Order(n^2)$ gates \cite{LaVe08}, the number of single- and two-qubit gates that suffice to implement $R$ scales as $\Order(nm)=\Order[n\log(n)]$. In order to see this, we determine the number of elementary gates of each factor of the decomposition of $R$ given in \eq{Rcircuit}.

As we will show in Appendix \ref{ap.appendixB}, for each $1\leq s \leq m-1$ the matrix $R_{S^{(s)}}$ is a single two-qubit gate. Given that $R_{S^{(0)}}$ can be decomposed as a product of $m-1$  $R_{S^{(s)}}$ matrices [see \eq{RS0} in Appendix \ref{ap.appendixB}], it follows that $R_{S^{(0)}}$ can be implemented with $m-1$ two-qubit gates. The gate $R_{S^{(m)}}$ can also be decomposed into a similar product of $\lfloor m/2\rfloor$ two-qubit gates [see \eq{RSm} in Appendix \ref{ap.appendixB}]. Hence, the implementation of $R_{S^{(k)}}$, for any $k$, requires at most $\Order(m)$ two-qubit gates. The matrix $R_{S^{\mathrm{Bog}}}$ can be implemented with $\Order(m^2)=\Order[\log(n)^2]$ elementary gates as can be seen as follows. Due to \eq{RSBog} (see Appendix \ref{ap.appendixB}) it is clear that $R_{S^{\mathrm{Bog}}}$ can be written as a product of three gates controlled by two qubits plus some small constant number of elementary gates. From the decomposition given in Appendix \ref{ap.appendixB} it can be seen that each of these gates require at most $\Order(m^2)$ elementary gates.

The implementation of $R_{F^{(s)}}$ can be done with $\Order(s)$ single- and two-qubits gates for $0\leq s\leq m-1$. In order to see that, first note that the matrices $R_G$ and $V_F$ appearing in the decomposition of $R_{F^{(s)}}$ given in \eq{RFsdecomposition} are single two-qubit gates. Additionally, using the expression of $\varphi(l,s)$ given in \eq{varphi}, the gate $D_{F^{(s)}}$, given in \eq{DFs}, can be written as a product of $2s$ single-qubit gates controlled by two qubits. Therefore, $D_{F^{(s)}}$ can be implemented with $\Order{(s)}$ elementary gates \cite{BaBe95}. It follows that $\Order(m^2)$ elementary gates are sufficient to implement all the compressed gates associated to the Fourier transformation.

The matrix $R_B$ can be implemented with $\Order(nm)$ single- and two-qubit gates, which is the leading scaling order in the number of elementary gates that decompose $R$. This scaling is due to the implementation of $D_B$ [see \eq{DB}]. One can easily see that $D_B$ can be decomposed into a product of $\frac{n}{2}-1$ single-qubit gates controlled by $m-1$ qubits, each of which can be implemented with $\Order(m)$ elementary gates \cite{BaBe95}. Note that due to the indirect dependance of the phases $\theta_l$ with respect to the index $l$, it seems unfeasible that $D_B$ can be implemented in a more efficient way. Hence, the compressed circuit can be implemented with $\Order(mn)$ elementary gates.

Let us point out that the implementation of the gates corresponding only to the Fourier transformation, i.e. the implementation of the product $R_{S^{(m)}}R_{F^{(m-1)}}R_{S^{(m-1)}}\dots R_{F^{(0)}}R_{S^{(0)}}$ can be done with $\Order\left[\log(n)^2\right]$ single- and two-qubit gates. In contrast to that, the number of nearest neighbor matchgates required to implement the Fourier transform in the original matchgate circuit scales as $\Order(n^2)$ \cite{LaVe08}.

In the subsequent section we are going to utilize the compressed gate $R$, given in \eq{Rcircuit}, to construct compressed circuits which can be employed to measure physically relevant quantities using exponentially less qubits.

\section{Applications}\label{sect.applications}

In \sect{decomposition} we recalled how to decompose the matrix $U$, which is diagonalizing the XY--Hamiltonian, into a product of matchgates. Any excited state or thermal state can be generated by applying $U$ to a product state, as we will recall in this section. Once this state is prepared, it is possible to measure observables whose expectation values reveal physically interesting properties of the system. The matchgate circuits consists of three steps: the preparation of a factorizable initial state, the evolution of the system by the unitary $U$, and the measurement of the desired observable. The aim of this section is to provide compressed quantum circuits that simulate these matchgates circuits. The condition that the given matchgate circuits are compressible in the way described in \sect{preliminaries} imposes some restrictions on the kind of observables that can be considered. Due to that, we consider here measurement operators that can be written as quadratic polynomials of the Majorana operators given in \eq{JordanWigner}.

Let us first consider the preparation of an arbitrary excited state of the XY Hamiltonian by applying $U$ to the eigenstates of $H[a]$. For convenience, let us combine the internal parameters $g$ and $\delta$ of the XY Hamiltonian $H$, given by \eq{Hamiltonian}, in one variable $\varphi\equiv (g,\delta)$. According to \eq{Ha} the eigenstates of the diagonal Hamiltonian $H(\varphi)[a]$, are the computational basis states $\ket{k}$ while its corresponding energies are given by
\beq
    \label{eq.Ek2}
    E_k(\varphi)=E_0+\sum_{l=0}^{n-1}\epsilon_l^{k_l}(\varphi),
\eeq
where $k_l$ are the binary components of $k$ and the energies $\epsilon_l$ are given in \eq{epsilons}. Let us denote by $\ket{\Psi_k(\varphi)}$ the eigenstate of $H(\varphi)[c]$ corresponding to $E_k(\varphi)$. As the product of matchgates $U(\varphi)$ presented in \sect{decomposition} maps $H(\varphi)[c]$ into the diagonal Hamiltonian $H(\varphi)[a]$ [see \eq{UHU}], we have
\beq
    \label{eq.psik}
    \ket{\Psi_k(\varphi)}=U(\varphi)\ket{k}.
\eeq

Thermal states of the Hamiltonian $H[c](\varphi)$ can also be generated by evolving the thermal state of the Hamiltonian $H[a]$ with the unitary $U(\varphi)$ \cite{LaVe08}. Note that the thermal state of the diagonal Hamiltonian $H(\varphi)[a]$ is given by $\rho(\varphi,T)[a]=\frac{1}{Z(\varphi)}\sum_{k}\epower^{-E_{k}(\varphi)/T}\proj{k}$, where $Z(\varphi)$ denotes the partition function $Z(\varphi)=\sum_{k}\epower^{-E_k(\varphi)/T}$. Due to the fact that
\beq
    \label{eq.Haexpansion}
    H(\varphi)[a]=-\frac{1}{2}\sum_{j=0}^{n-1}{\epsilon_j Z_j},
\eeq
the state $\rho(\varphi,T)[a]$ can be factorized as
\beq
    \label{eq.rhoa}
    \bal
        \rho(\varphi,T)[a]&\equiv\bigotimes_{k=0}^{n-1}\rho_k(\varphi,T)[a].
    \eal
\eeq
The local density matrices $\rho_k(\varphi,T)$ are diagonal and are given by
\beq
    \label{eq.rhogTproduct}
    \rho_k(\varphi,T)[a]=a_k(\varphi,T)\proj{0}+b_k(\varphi,T)\proj{1},
\eeq
where $a_k(\varphi,T)=(1+\epower^{-\frac{\epsilon_k(\varphi)}{T}})^{-1}$ and $b_k(\varphi,T)=\epower^{-\frac{\epsilon_k(\varphi)}{T}}a_k(\varphi,T)$. The thermal state of the Hamiltonian $H(\varphi)[c]$ is obtained by evolving $\rho(\varphi,T)[a]$ under $U(\varphi)$, that is
\beq
    \rho(\varphi,T)[c]=U(\varphi)\rho(\varphi,T)[a]\adj{U}(\varphi).
\eeq

Thus, thermal states and any eigenstates of the XY Hamiltonian can be prepared evolving the system that is initially prepared in a product state with the unitary $U(\varphi)$. The final element of the circuit is a measurement of some observable on these states. As mentioned above, in order to be able to compress the matchgate circuit, the measurement operators are chosen as arbitrary operators $A$, acting on $n$ qubits, which can be written in term of the Majorana operators $x_j$ given in \eq{JordanWigner} as
\beq
    \label{eq.Aquadratic}
    A=\sum_{j\neq k=0}^{2n-1}a_{j,k}x_jx_k.
\eeq

In the following subsections we present the compressed quantum circuits running on $m+1=\log(n)+1$ qubits, corresponding to matchgate circuits whose outputs are the expectation values $\expect{A}(\varphi,T)\equiv\tr\left\{\rho(\varphi,T)[c]A\right\}$ or $\expect{A}(\varphi,k)\equiv\bmk{\Psi_k(\varphi)}{A}{\Psi_k(\varphi)}$, where $\rho(\varphi,T)$ ($\ket{\Psi_k(\varphi)}$) denotes the Gibbs state (eigenstate) of the XY Hamiltonian of $n$ qubits respectively. We will then consider three examples of physically-relevant measurement operators that can be written in the form given in \eq{Aquadratic}, and will demonstrate how time evolution can be studied with an exponentially smaller system.

\subsection{Construction of the compressed circuit for thermal states}\label{sect.compressedGibbs}

Here, we consider the matchgate circuit that leads to the measurement of the observable $A$ on the thermal state at temperature $T$. As shown above, the outcome of this circuit can be written as $\expect{A}(\varphi,T)=\tr\left\{U(\varphi)\rho(\varphi,T)[a]\adj{U}(\varphi)A\right\}$. Using \eq{UxU} and the explicit expression of the observable $A$ given in \eq{Aquadratic} we have
\beq
    \label{eq.expectA1}
    \bal
        \expect{A}(\varphi,T)&=\sum_{j,k=0}^{2n-1}a_{j,k}\bmk{j}{R(\varphi)S(\varphi,T)R^{T}(\varphi)}{k},
    \eal
\eeq
where the $2n\times 2n$ matrix $S(\varphi,T)$ is defined as $\left[S(\varphi,T)\right]_{r,s}\equiv\tr\left\{-ix_rx_s\rho(\varphi,T)[a]\right\}$. Using the expression for $\rho(\varphi,T)[a]$ given in \eq{rhogTproduct}, one can see that the entries, $[S(\varphi,T)]_{r,s}$ of $S(\varphi,T)$ vanish unless $(r,s)=(2l,2l+1)$ or $(r,s)=(2l+1,2l)$, for some $l$. Explicitly, we find that
\beq
    S(\varphi,T)=\sum_{l=0}^{n-1}\tanh\left[\frac{\epsilon_l(\varphi)}{2T}\right]\big(\kb{2l}{2l+1}-\kb{2l+1}{2l}\big).
\eeq
Let us define the positive operator $\bar{\rho}(\varphi,T)[a]=\frac{1}{2n}\left[\one-iS(\varphi,T)\right]$ acting on $m+1$ qubits. As we will see this operator corresponds to the initial state of the compressed circuit running on $m+1$ qubits. One can easily verify that this state can also be written as
\beq
    \label{eq.barrho}
    \bal
        \bar{\rho}(\varphi,T)[a]&=\frac{1}{n}\sum_{l=0}^{n-1}\proj{l}\otimes\Big[a_l(\varphi,T)\proj{+_y}_m\\
        &\phantom{=}+b_l(\varphi,T)\proj{-_y}_m\Big],
    \eal
\eeq
where $\ket{+_y}$ and $\ket{-_y}$ are the eigenstates of $Y$ corresponding to the eigenvalues $1$ and $-1$ respectively. Using \eq{expectA1} and the fact that $R$ is orthogonal, the expectation value $\expect{A}(\varphi,T)$ can now be written as
\beq
    \label{eq.expectAcompressed}
    \expect{A}(\varphi,T)=n\tr\left[R(\varphi)\bar{\rho}(\varphi,T)[a] R^T(\varphi)\overline{A}\right].
\eeq
Here, we have defined the observable $\overline{A}$ acting on the Hilbert space of $m+1$ qubits as
\beq
    \label{eq.overlineA}
    \overline{A}=i\sum_{j,k=0}^{2n-1}a_{j,k}\left(\kb{j}{k}-\kb{k}{j}\right).
\eeq
With all that, the expectation value of any observable $A$ [see \eq{Aquadratic}] can be measured using the following compressed circuit:
\begin{enumerate}[(i)]
    \item prepare the initial $(m+1)$--qubit state $\bar{\rho}(\varphi,T)[a]$, given in \eq{barrho},
    \item evolve the state according to the orthogonal matrix $R(\varphi)$ given in \eq{Rcircuit},
    \item measure the operator $\overline{A}$, given in \eq{overlineA}.
\end{enumerate}
Below, we consider three particular examples of physically relevant quantities which can be measured with observables that are of the form given in \eq{Aquadratic} and therefore can be measured with a circuit as the one described above. Note that $\bar{A}$ can always be mapped to a local observable requiring at most $\Order(n)$ elementary gates.

\subsubsection{Magnetization}\label{sect.aplicationmagnetization}

We discussed in \sect{preliminaries} that the XY Hamiltonian exhibits a phase transition as a function of the parameter $g$ [cf. \eq{Hamiltonian}]. As the magnetization shows an abrupt behavior at the critical value, $g_c$, it can be utilized to observe this phase transition. In \cite{Kr10,BoMu12} we derived compressed circuits to measure the magnetization of the Ising and XY model by simulating an adiabatic evolution. With such a circuit, the magnetization of the groundstate for any value of $g$ can be measured. Here, we derive a compressed circuit to measure the magnetization of both thermal states and arbitrary excited states of the XY Hamiltonian. In contrast to \cite{Kr10, BoMu12} we will not simulate here the adiabatic evolution, but simulate the evolution of $U(\varphi)$.

The magnetization is the expectation value of an observable $M=\frac{1}{n}\sum_j Z_j$ which can be written as
\beq
    M=-\frac{i}{n}\sum_{j=0}^{n-1}x_{2j}x_{2j+1}.
\eeq
Using this expression for $M$ and \eq{overlineA}, one can easily compute the operator $\overline{M}$ which needs to be measured in the compressed circuit. It is given by
\beq
    \label{eq.barM}
    \overline{M}=-\frac{i}{n}\sum_{j=0}^{n-1}\big(\kb{2j}{2j+1}-\kb{2j+1}{2j}\big)=\frac{1}{n}\left(\one\otimes Y_m\right),
\eeq
where $Y_m$ is the Pauli operator acting on the $m^{th}$ qubit. Using now \eq{expectAcompressed} we have that
\beq
    \label{eq.expectMcompressed}
    \expect{M}(\varphi,T)=\tr\left[R(\varphi)\bar{\rho}(\varphi,T)R^T(\varphi)(\one\otimes Y_m)\right].
\eeq

Hence, the magnetization of the thermal state of the XY Hamiltonian, $\expect{M}(\varphi,T)$, can be measured with the following compressed circuit running on $\log(n)+1$ qubits. First the initial $m+1$ qubit state $\bar{\rho}(\varphi,T)$ for the desired temperature [see \eq{barrho}] is prepared. The gate $R$, decomposed as in \eq{Rcircuit}, is applied and finally the last qubit is measured in $y$--direction.

\begin{figure}[h]
\centering
    \includegraphics[width=0.23\textwidth]{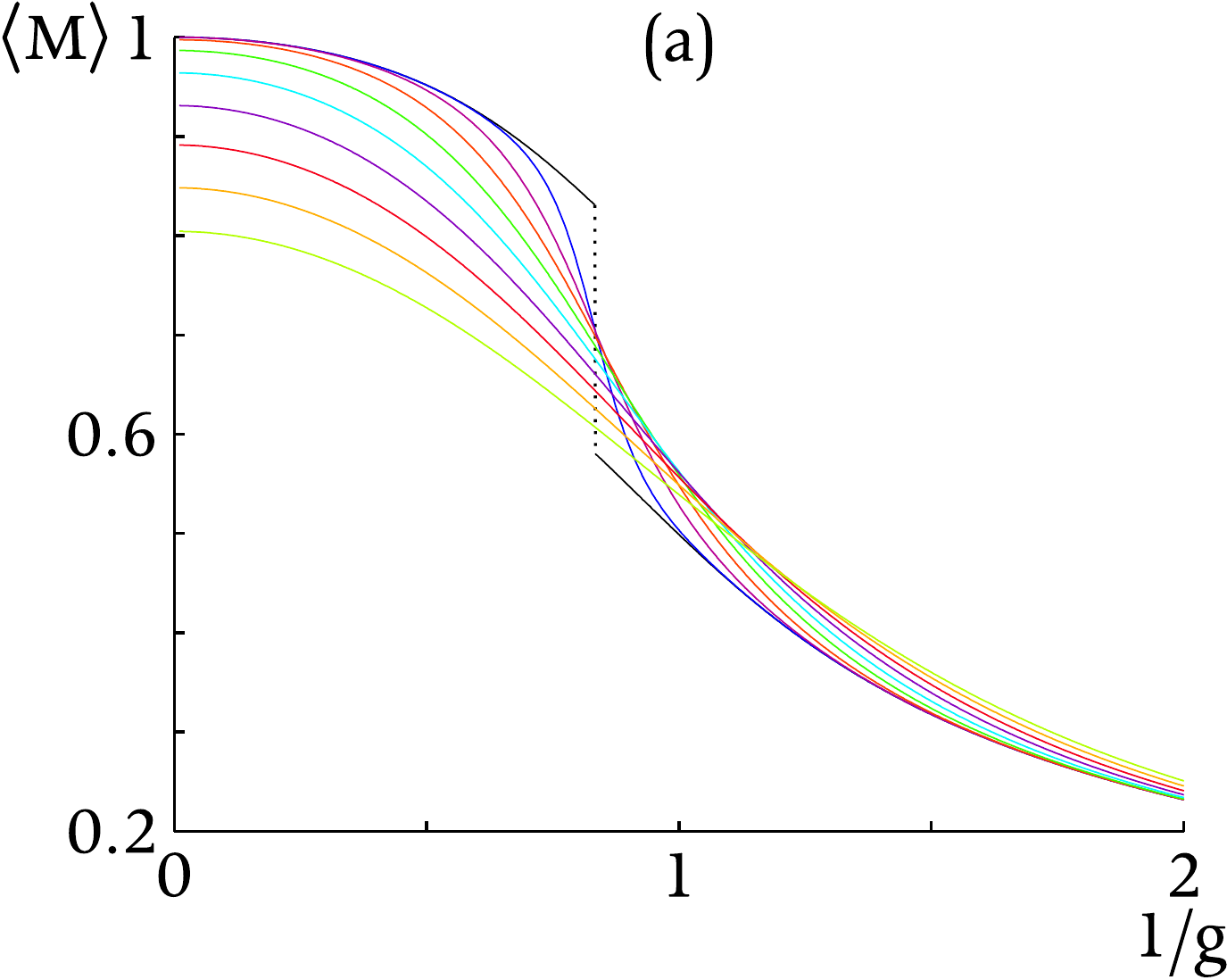}
    \includegraphics[width=0.23\textwidth]{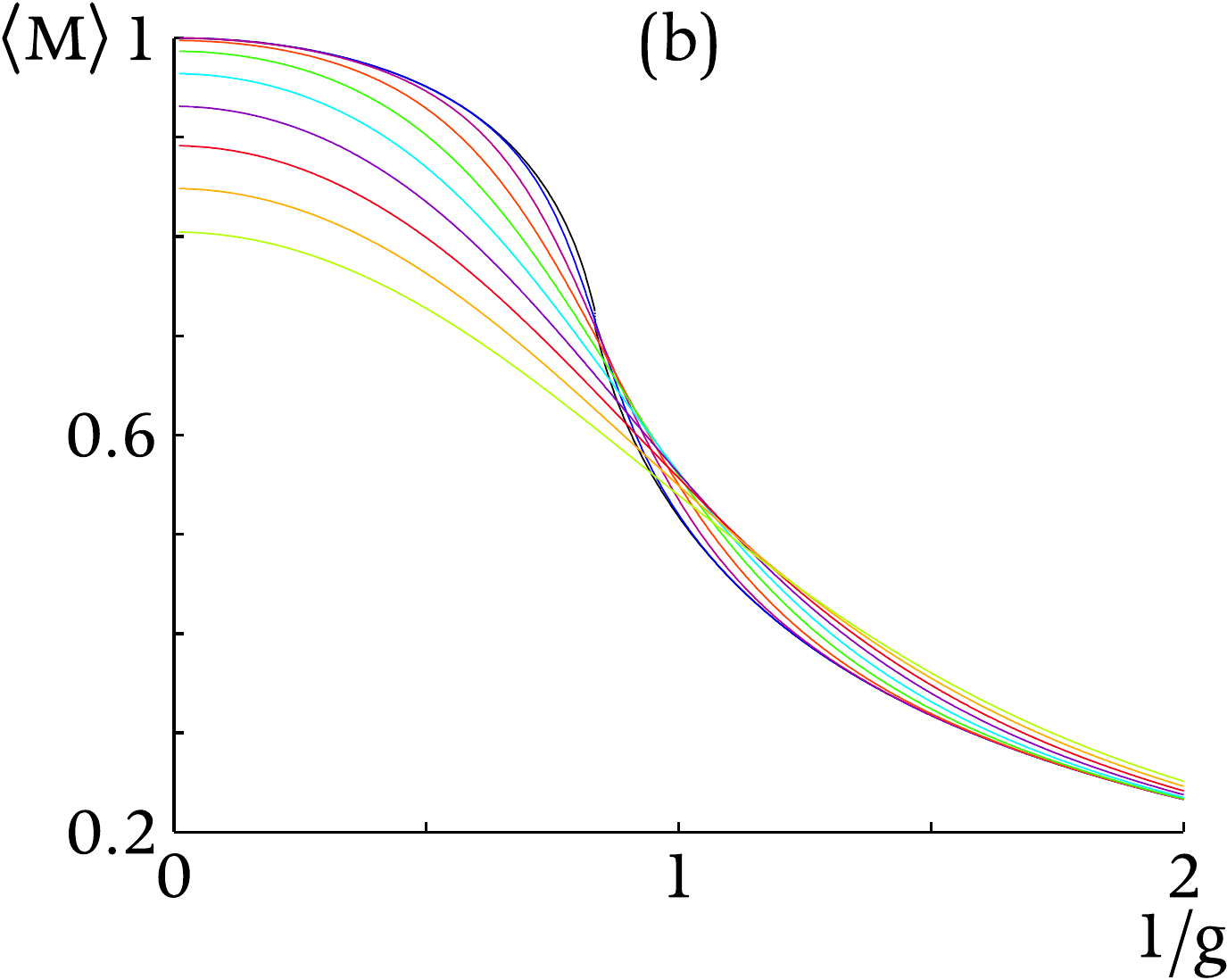}
    \caption{(Color online) The magnetization [see \eq{expectMcompressed}] of the thermal states of the XY Hamiltonian as a function of $g$ is shown. We have set $\delta=0.2$ and considered system sizes of $n=8$ qubits (a) and $n=128$ qubits (b). Each curve corresponds to a different value of the temperature, which is chosen equally spaced between $T=0$ (uppermost curve on the left side of the figures) and $T=0.9$ (lowermost curve on the left side of the figures). At $T=0$ the magnetization shows a discontinuity at the critical point (dashed line) as a consequence of the level crossing between the groundstate and the first-exited state. The discontinuity diminishes as the system size increases [compare figures (a) and (b)] and vanishes if $T>0$.\label{fig.magnetization}}
\end{figure}

\figu{magnetization} shows the magnetization of the thermal states of the XY Hamiltonian as a function of $g$. The curves plotted correspond to two different system sizes, $n=8$ in \figu{magnetization} (a) and $n=128$ in \figu{magnetization} (b). We have also plotted the magnetization corresponding to thermal states at temperatures between $T=0$ and $T=0.9$. In case $T=0$, one observes that the magnetization exhibits a discontinuity at the critical point $g_c$. This behaviour is due to a level crossing occurring between the groundstate and the first excited state, when Jordan-Wigner boundary conditions are considered (see \sect{preliminaries}). This sudden change in the groundstate of the system, as a function of $g$, is reflected as a discontinuity of the magnetization of the system. Given that this behavior is due to the boundary conditions, it is a finite size effect and becomes negligible as the system size grows. This can be seen when comparing figures (a) and (b). Note also that the temperature removes the discontinuity of the magnetization for all of the considered temperatures. This is expected as for any $T>0$ and any value of $g$, the Gibbs state presents a mixture of the ground and excited states with some nonzero Boltzmann weights. Let us remark that this discontinuity is not due to the phase transition. In fact, if one considers only eigenstates with a given parity and momentum, the magnetization of the groundstate at $T=0$ remains continuous at the critical point (see also \cite{BoMu12} and references therein) and the phase transition can be witnessed as a discontinuity of the second derivative of the magnetization.

\subsubsection{Quantum Quenching}\label{sect.aplicationquenching}

In \cite{BoMu12} we derived a matchgate circuit and its corresponding compressed circuit to measure the number of kinks generated in a chain of Ising qubits, after the parameter $g$ is quenched from a large value, $\gmax$, down to $0$. The number of kinks quantifies the number of spins that are not aligned in the $X$ direction with respect to their neighboring spins. This number, normalized by the number of spins, can be quantified by $\nu=\frac{1}{2}(1-\expect{K})$ \cite{BoMu12}, where the observable $K$ is defined as
\beq
    \label{eq.K}
    \bal
        K&=\frac{1}{n-1}\sum_{j=0}^{n-2}X_jX_{j+1}\\
        &=-\frac{i}{n-1}\sum_{j=0}^{n-2}x_{2j+1}x_{2j+2}.
    \eal
\eeq

We present now a circuit which can be utilized to measure $\nu$ as a function of the quenching time $t$, i.e., the time used to quench the parameter $g$, for thermal and excited states. As mentioned before, $\nu$ corresponds to the number of kinks only for the groundstate. However, as $\nu$ can be also viewed as a measure of the correlations in the system, we investigate here its value for thermal as well as excited states.

For convenience, we separate the circuit into two parts. In the first part the Gibbs state of the Hamiltonian $H(\gmax)[c]$, $\rho(\gmax,T)[c]$, is prepared. This is accomplished by initially preparing the system in the state $\rho(\gmax,T)[a]$ given by \eq{rhoa}, and evolving it with the unitary $U(\gmax)$ presented in \sect{decomposition}. The corresponding compressed initial state and orthogonal gate are $\bar{\rho}(\gmax,T)$, given in \eq{barrho} and $R(\gmax)$ [see \eq{Rcircuit}] respectively.

The second part of the evolution corresponds to the quantum quenching. In this stage of the circuit, a product of matchgates $U_Q(t)$ simulates the quenching of the parameter $g$ from $\gmax$ down to $0$. We have provided the derivation of the unitary $U_Q(t)$, and its corresponding compressed gate $R_Q(t)$ in \cite{BoMu12}. Here, we simply recall the final decomposition of the matrix $R_Q(t)$. The unitary $U_Q(t)$ consists of a discrete product of Trotter steps, approximating the continuous change of the parameter $g$. Denoting the number of steps by $L+1$, the error accumulated by this approximation scales as $\Order(L\Delta^2)$, with $\Delta(t)=\frac{t}{L+1}$. The compressed gate $R_Q(t)$ corresponding to the product of matchgates $U_Q(t)$, is given by
\beq
    \label{eq.RQ}
    R_Q(t)=\prod_{l=0}^{L}R_0(t,l)R_1(t),
\eeq
where
\beq
    \bal
        R_0(t,l)&=\cos\left[2g_l\Delta(t)\right]\one+\sin\left[2g_l\Delta(t)\right]h_0\\
        R_1(t)&=\cos\left[2\Delta(t)\right]\one+\sin\left[2\Delta(t)\right]h_1,\\
    \eal
\eeq
with $g_l=\gmax\left(1-\frac{l}{L}\right)$, and
\beq
    \bal
        h_0&=\sum_{j=0}^{n-1}\left(\kb{2j+1}{2j}-\mathrm{h.c.}\right)\\
        h_1&=\sum_{j=1}^{n-1}\left(\kb{2j+2}{2j+1}-\mathrm{h.c.}\right).\\
    \eal
\eeq

Using \eq{overlineA}, one can see that the observable $\overline{K}$ that has to be measured in the compressed circuit is given by
\beq
    \label{eq.barK}
    \overline{K}=-\frac{i}{n-1}\sum_{j=0}^{n-2}\big(\kb{2j+1}{2j+2}-\kb{2j+2}{2j+1}\big).
\eeq
With all that, it is straight forward to write the expectation value $\expect{K}(t)$ as the outcome of a compressed quantum circuit like the one given above [see \eq{expectAcompressed}]. We obtain
\beq
    \label{eq.expectK}
    \expect{K}(t)=\tr\left[R_Q(t)R(\gmax)\bar{\rho}(\gmax,T)[a]R^T(\gmax)R^T_Q(t)\overline{K}\right].
\eeq

Hence, the expectation value of the operator $K$ can be measured as follows:
\begin{enumerate}[(i)]
    \item prepare the initial $m+1$--qubit state $\bar{\rho}(g_{\mathrm{max}},T)[a]$, given in \eq{barrho},
    \item evolve the state according to the orthogonal matrix $R_{Q}(t)R_(\gmax)$ given by \eq{RQ} and \eq{Rcircuit},
    \item measure the operator $\overline{K}$, given in \eq{barK}.
\end{enumerate}

\begin{figure}[h]
\centering
    \begin{tabular}{cc}
        \includegraphics[width=0.20\textwidth]{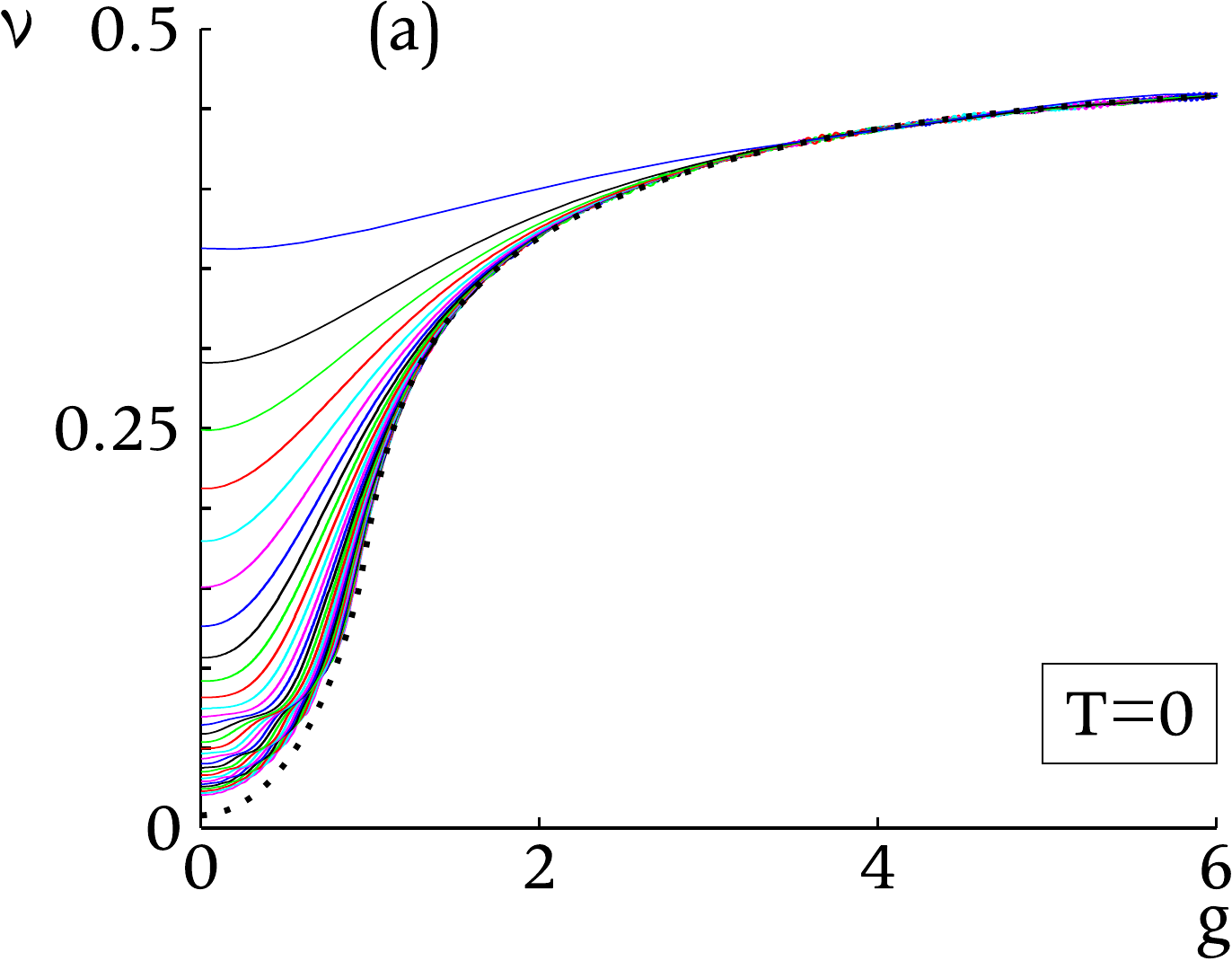} & \includegraphics[width=0.20\textwidth]{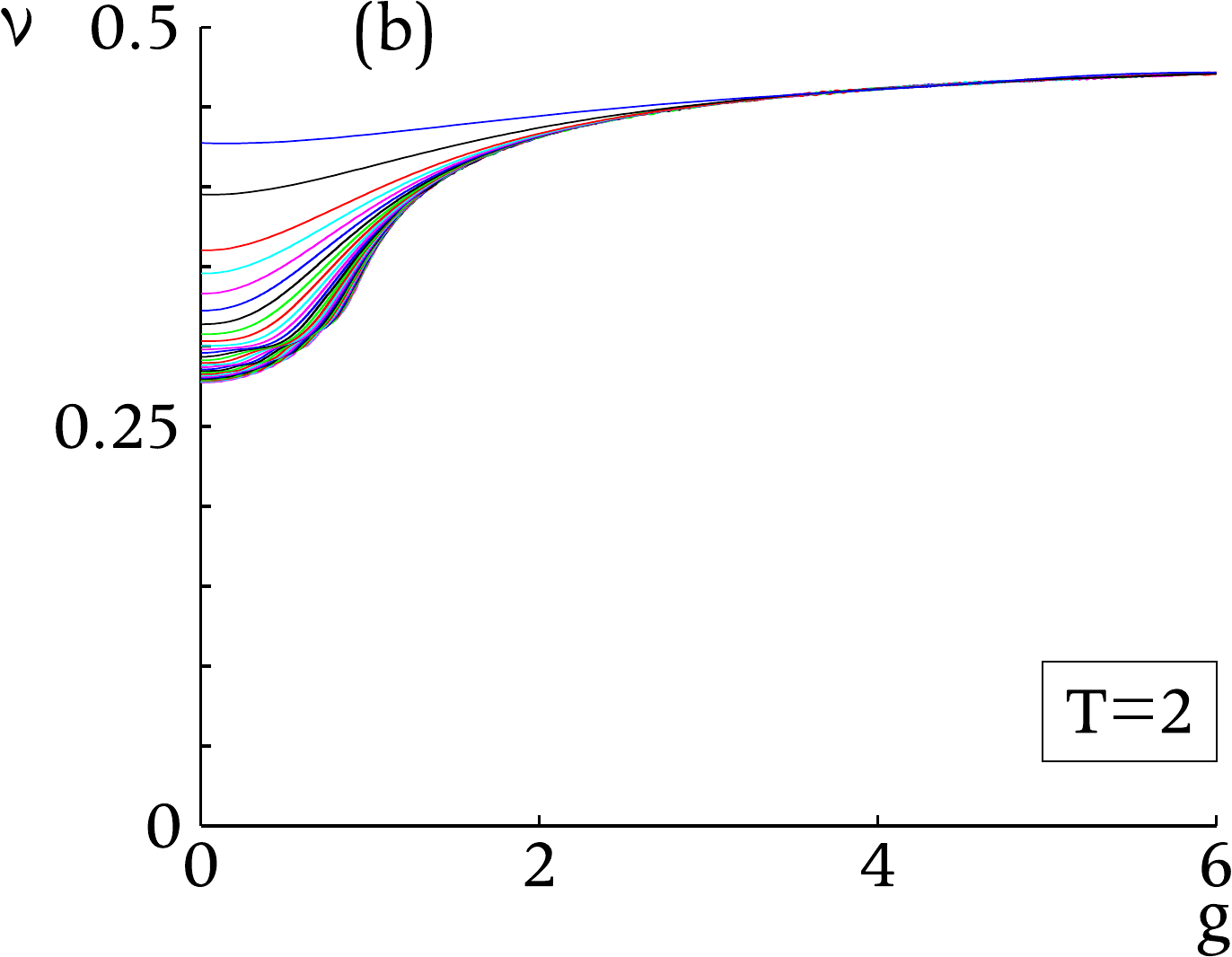} \\
        \includegraphics[width=0.23\textwidth]{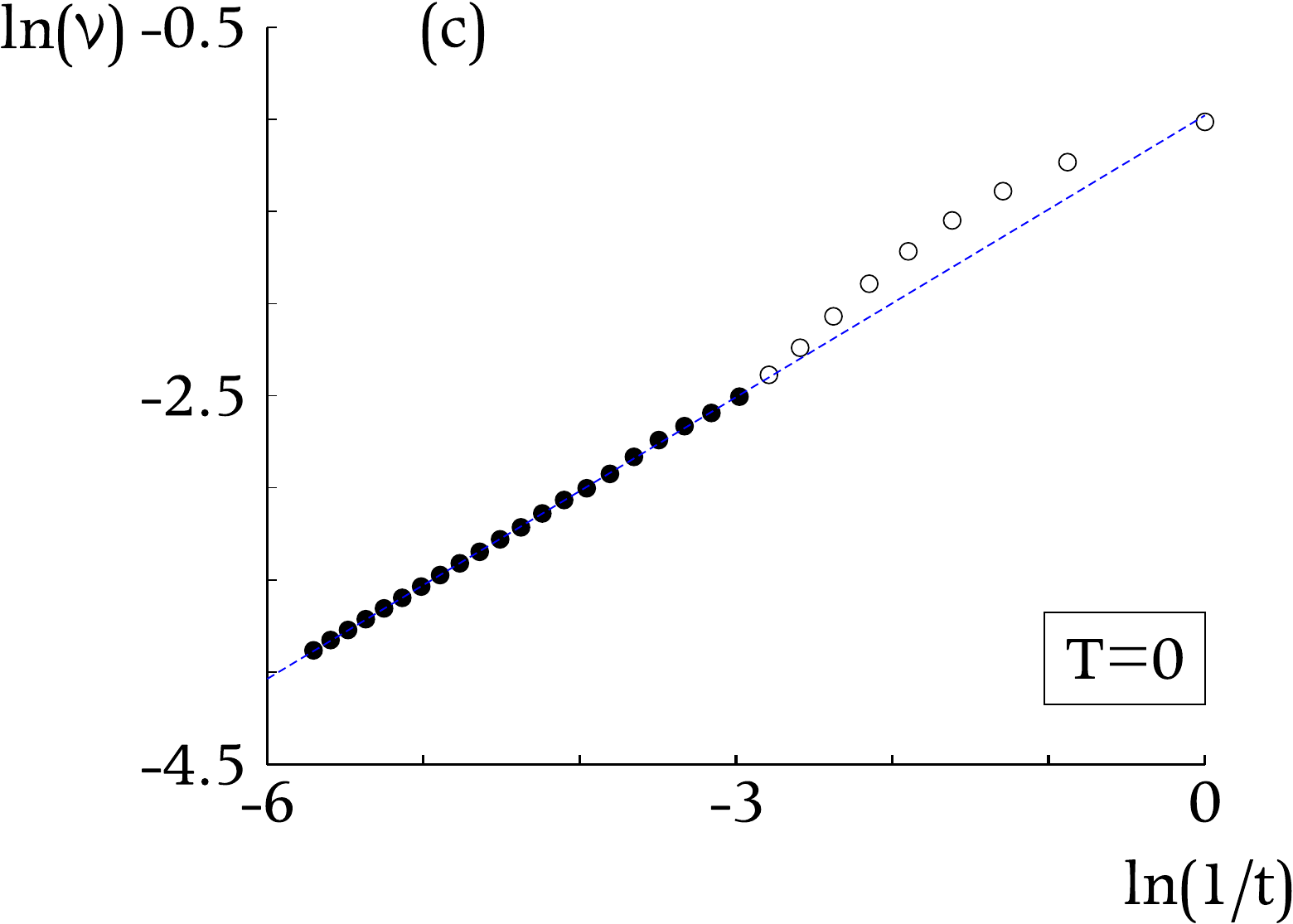} & \includegraphics[width=0.23\textwidth]{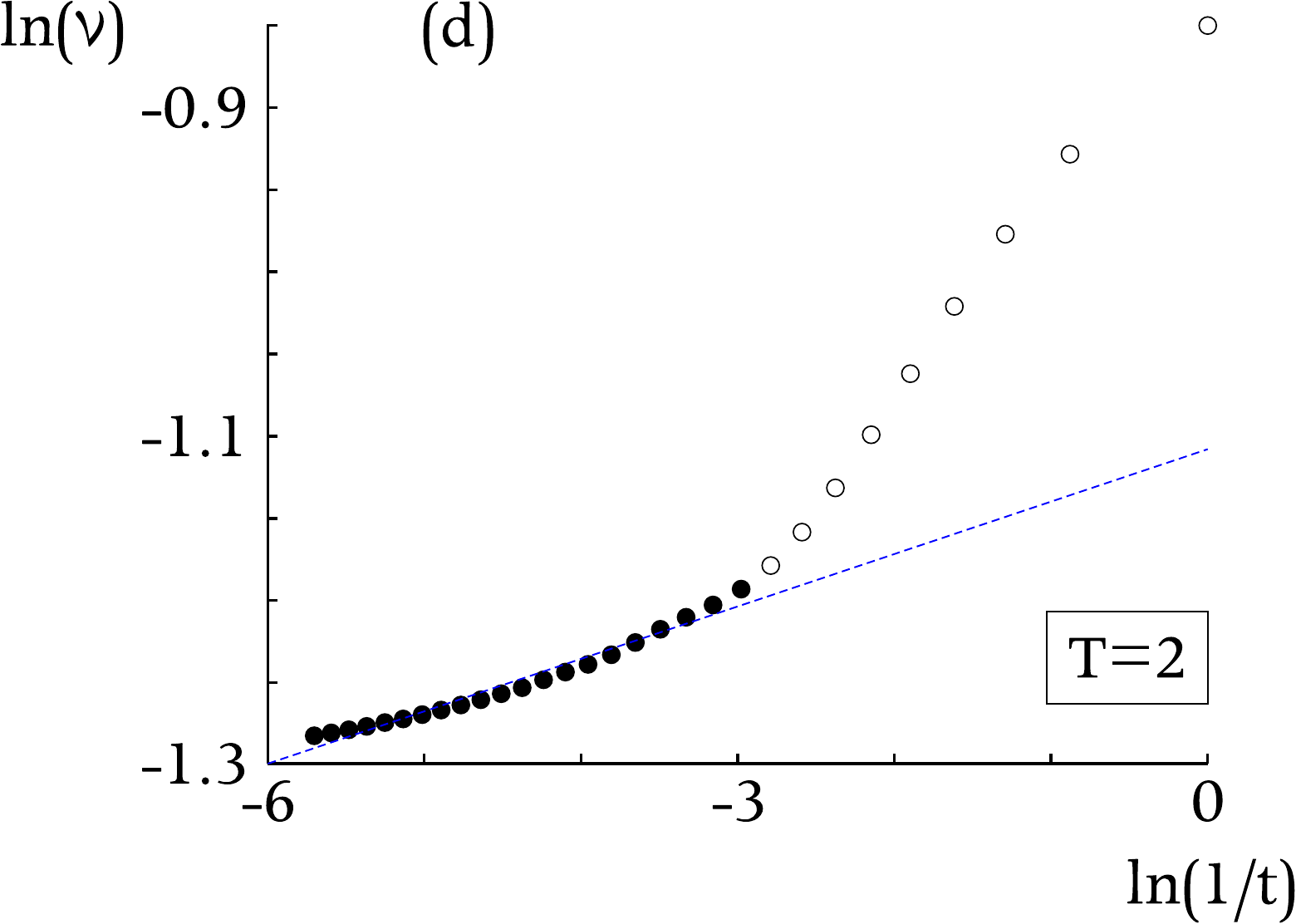} \\
    \end{tabular}
    \caption{(Color online.) The number of kinks, $\nu$, as a function of the parameter $g$, occurring during the quantum quenching [see \eq{expectK}] for a system size of $n=128$ spins is depicted. The different curves correspond to various total quenching times ranging from $t=1$ (uppermost curve) to $t=300$ (continuous bottommost curve). The number of Trotter steps was set to $L=100t$. The number of kinks for $T=0$ and adiabatic evolution is depicted in (a) as a dashed line.
    Figures (c) and (d) show a standard linear fitting function of the logarithm of the correlations $\nu$ at the end of the quenching, i.e. at $g=0$, vs the logarithm of the quenching speed ($1/t$). To fit the data points we have only considered the points depicted by filled circles, which correspond to long quenching times, although not long enough to correspond to an adiabatic evolution. The slope of the fitted function (represented by the dashed line) is the parameter $p$, corresponding to the exponent in the relation $\nu\propto t^{-p}$. We have obtained a value $p=0.51$ for $T=0$, and $p=0.03$ for $T=2$. \label{fig.kinks}}
\end{figure}

In \figu{kinks} (a) and (b), we depict the correlations $\nu$, computed numerically using \eq{expectK} as a function of $g$ during the quenching for different total quenching times $t$. We have considered a value $\gmax=10$. Each curve depicted in the figures corresponds to a different total quenching time, chosen between $t_{\mathrm{min}}=1$ and $t_{\mathrm{max}}=300$, given in absolute units. The uppermost curve in both figures corresponds to a short quenching time, $t=1$, while the bottommost continuous curve corresponds to the larger time $t=300$. For comparison we have used the matrix $R(g)$ to compute the correlations $\nu$ (for $T=0$) that would occur in the system if the evolution were to be performed adiabatically. The curve computed in this way is plotted with a dashed line in \figu{kinks} (a).

\begin{figure}[h]
\centering
    \includegraphics[width=0.4\textwidth]{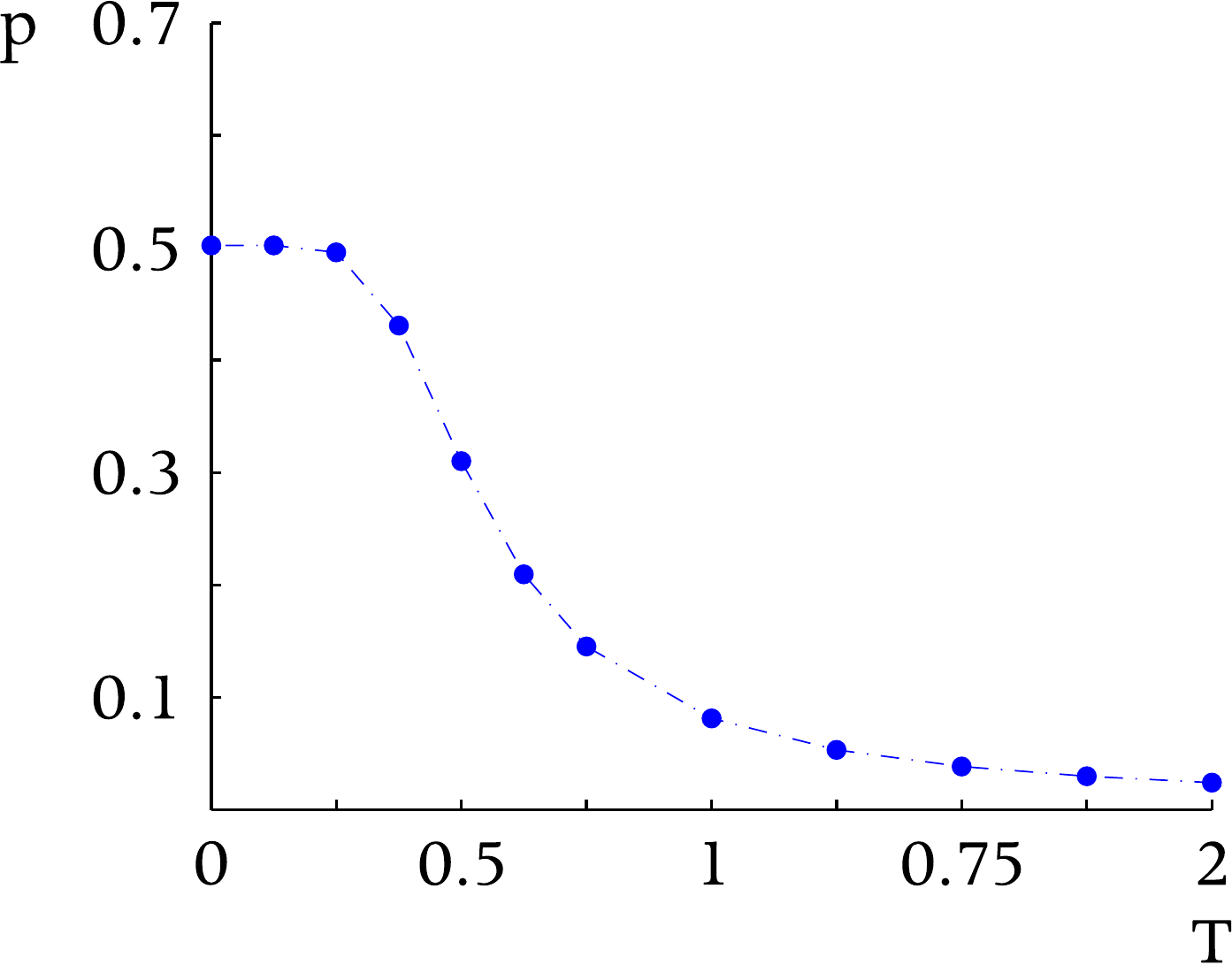}
    \caption{(Color online.) The parameter $p$ that corresponds to the exponent in the relation $\nu\propto t^{-p}$ is depicted as a function of the temperature $T$. Each point has been obtained by simulating a quantum quenching for a system of $n=64$ qubits. For each value of the temperature we have extracted $p$ by fitting the correlations $\nu$ measured in the system after the quantum quenching as a function of the quenching time $t$ (see \figu{kinks}). Note that for low temperatures, the parameter $p$ is close to the one for $T=0$. This is expected as the correlations $\nu$ in each of the first four excited states scale similarly as the ones of the ground state. \label{fig.kinkstemperature}}
\end{figure}

\begin{figure}[h]
\centering
    \begin{tabular}{cc}
        \includegraphics[width=0.23\textwidth]{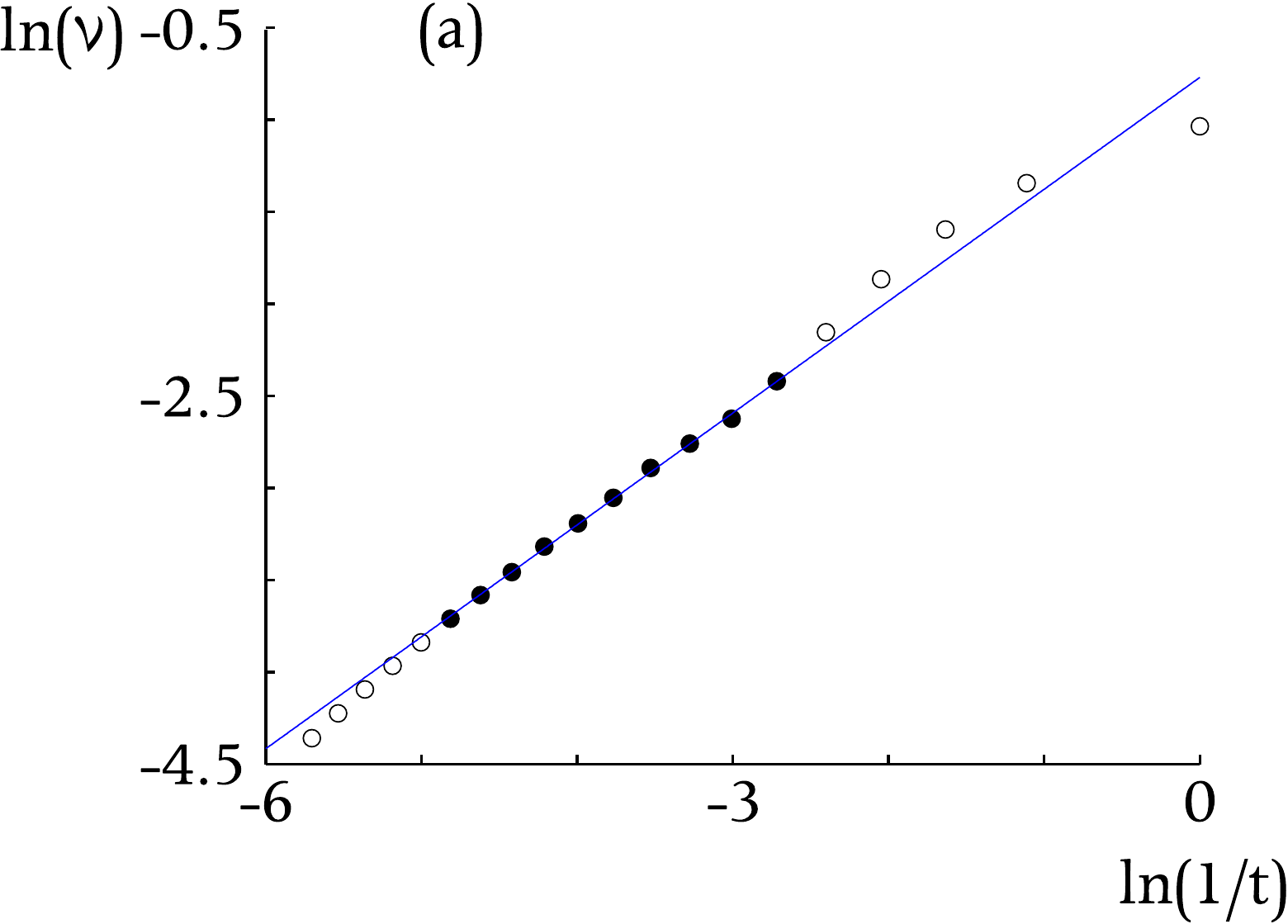} & \includegraphics[width=0.23\textwidth]{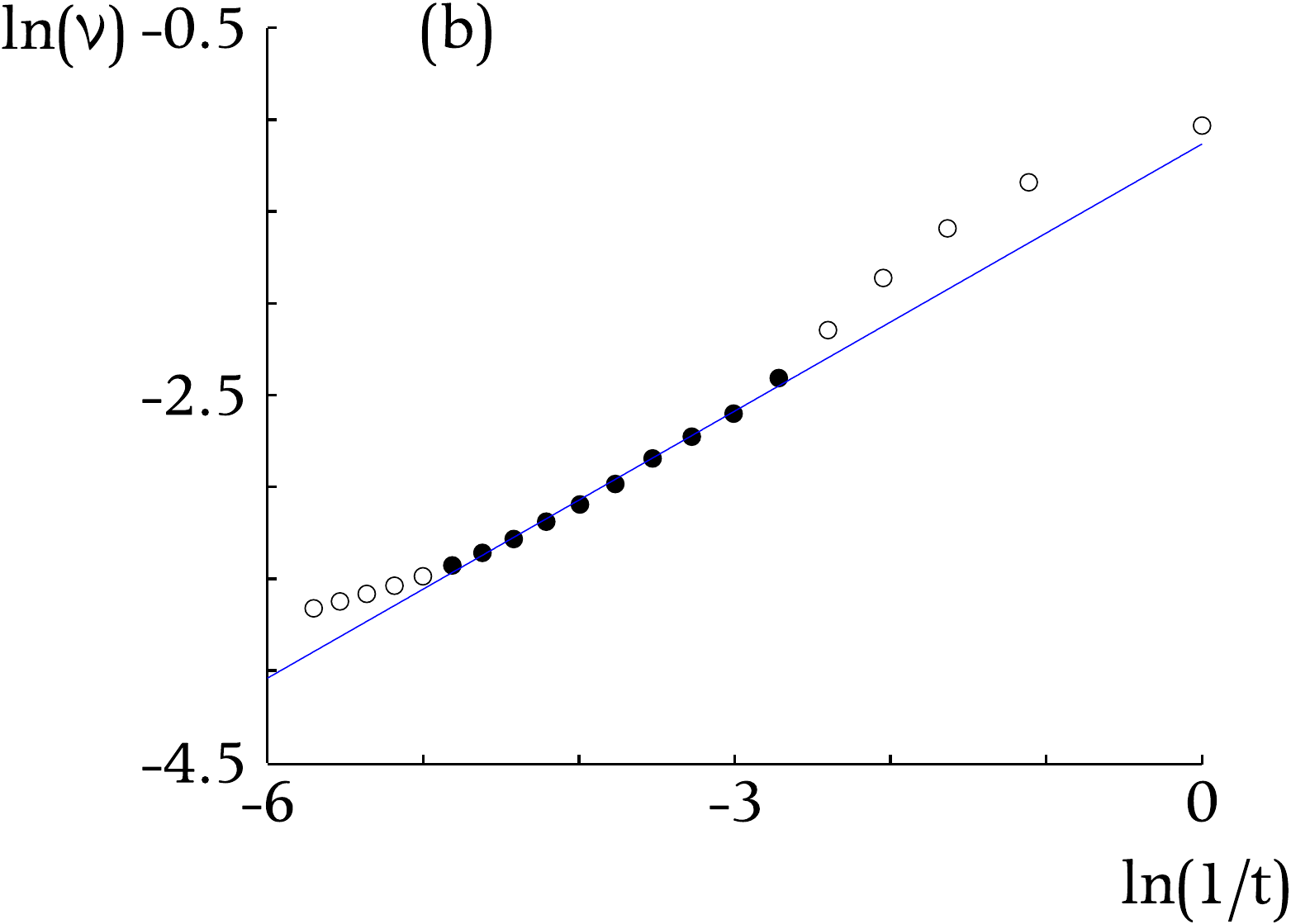} \\
    \end{tabular}
    \caption{(Color online.) The linear fit of the logarithm of the correlations $\nu$ at the end of the quenching, i.e. at $g=0$, vs the logarithm of the quenching speed ($1/t$) is shown. The simulated system size is of $n=128$ qubits. The fitting was done in the same way as explained in \figu{kinks}. To obtain the correlations depicted in (a) and (b) we have considered a system initially prepared in the first and fourth excited state of the Ising Hamiltonian respectively. The corresponding slopes are $p=0.61$ (a) and $p=0.48$ (b). \label{fig.kinksexcited}}
\end{figure}

The Kibble-Zureck mechanism establishes at $T=0$ a relation $\nu(t)\propto (1/t)^p$, with $p=1/2$, between the number of kinks and the quenching speed $(1/t)$. We use the data plotted in \figu{kinks} (a) to verify that such a mechanism is present in the Ising chain at $T=0$ (see also \cite{BoMu12}). In order to do so, we plot the logarithm of the correlations $\nu$ at the end of the quenching evolution, i.e. at $g=0$ as a function of the logarithm of the quenching speed. Due to the Kibble-Zureck mechanism, we expect a linear behavior of this function, with a slope $p=1/2$. The data points have been fitted by a linear function in the region where a linear behaviour is observed, yielding $p=0.51$ for a system size of $n=128$ qubits [see \figu{kinks} (c)], in agreement with the Kibble-Zureck mechanism. The case where $T>0$ lies out of the scope of the Kibble-Zureck mechanism and the relationship between the correlations $\nu$ (which at $T=0$ correspond to the number of kinks) and the quenching speed is up to our knowledge unknown. In order to obtain some insight in the case where $T>0$, we used \eq{expectK} to compute numerically, for various temperatures, the correlations $\nu$ present in a system initially prepared in a thermal state. As an example the values of $\ln{(\nu)}$ computed in the case where $T=2$ are plotted in \figu{kinks} (d) as a function of $\ln(1/t)$. In order to extract the proportionality factor $p$, we linearly fitted this function using the same interval of data points used for the case where $T=0$ [see \figu{kinks} (c)] and found that $p=0.03$. In \figu{kinkstemperature} we plot the value of $p$ obtained in the same way when considering temperatures $T \in [0,2]$. For low temperatures, there is a regime ($T\lesssim 0.25$) where the system still responds according to the Kibble-Zureck mechanism, as can be seen in \figu{kinkstemperature}. This behaviour can be understood as follows. For small enough temperatures only the groundstate and the first excited states have a non-negligible Boltzmann weight in the thermal state. Therefore, the evolution of the thermal state during the quenching evolution is determined by the evolution of the groundstate and the first exited states. The compressed circuit presented in this section can be easily adapted to simulate a quantum quenching in systems prepared initially in any eigenstate of the Hamiltonian
\footnote{
    We use the circuit that we present in \sect{compressedexcited} to simulate the preparation of an excited eigenstate as the initial state of the circuit. The part of the circuit that simulates the quenching and the final measurement are left unchanged.
}. Doing so, we computed the ratio $p$ between $\ln(\nu)$ and $\ln(1/t)$ for each of the first four excited states and found out that each of these values lie in the interval $[0.48,0.61]$ (for an example see \figu{kinksexcited} where we depict the logarithm of the correlations $\nu$ present in a system initially prepared in the first and fourth excited states). Hence, to a good approximation, the dependence of the correlations $\nu$ with respect to the quenching speed agrees with the Kibble-Zureck mechanism for low temperatures. However one can see that, as expected, the dependency of $\nu$ on the quenching speed is drastically changed for higher temperatures (see \figu{kinkstemperature}).


\subsubsection{Correlations}

\begin{figure}[h]
\centering
    \includegraphics[width=0.23\textwidth]{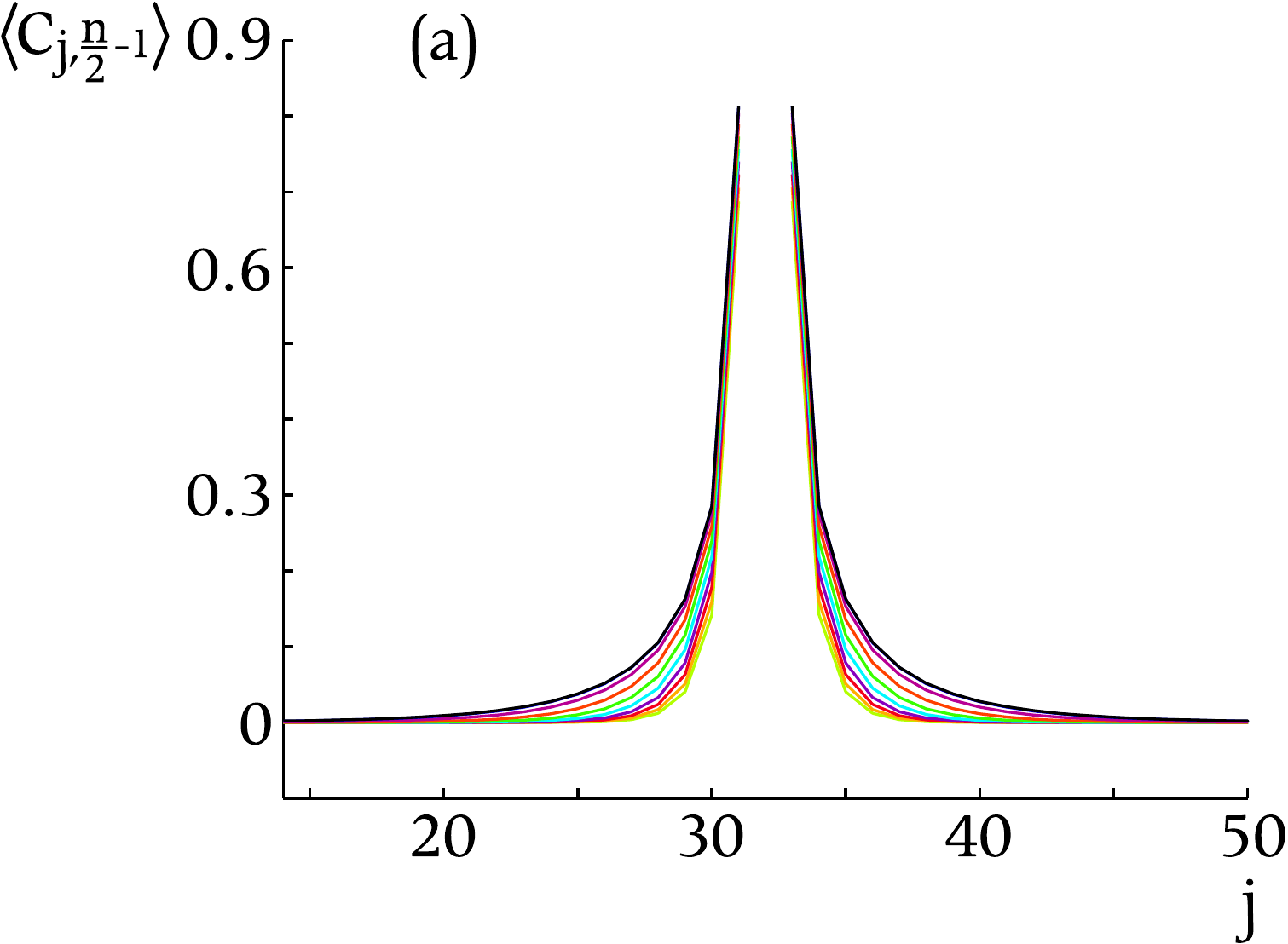}
    \includegraphics[width=0.23\textwidth]{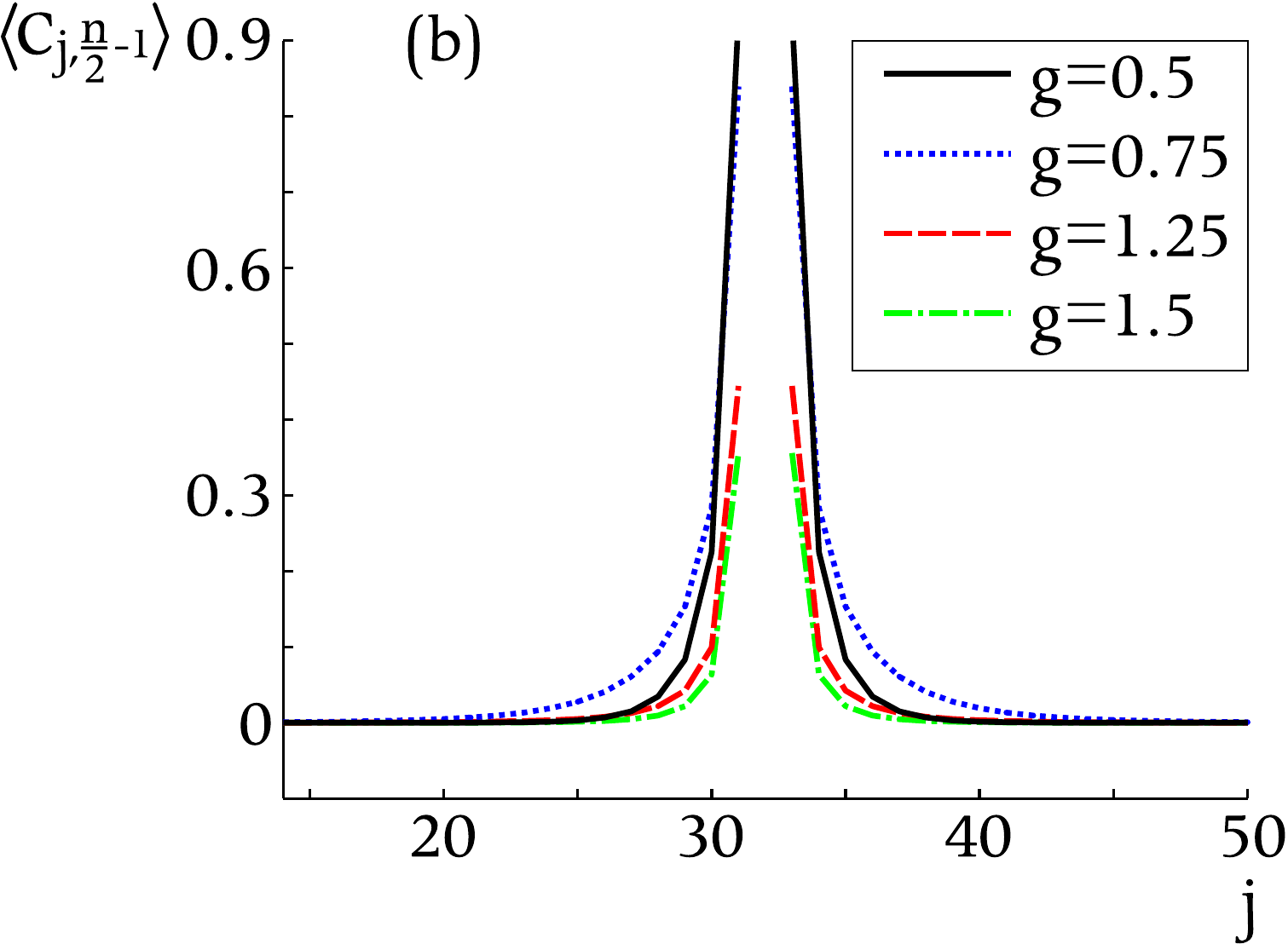}
    \caption{(Color online.) The correlations of the form $C_{j,k}$ [see \eq{Cjk}] between the $j$th qubit and the $(\frac{n}{2})$th qubit as a function of $j$ are depicted for $n=64$ qubits (not all depicted). In (a) the dependency of the correlations on temperature is shown. Each curve corresponds to a different temperature whose values have been chosen to be equally spaced between $T=0$ (corresponding to the uppermost curve) and $T=0.9$ (corresponding to the bottommost curve); the value of $g$ has been fixed at $g=0.8$. In (b) we show the dependency of the correlations on the value of $g$, at $T=0$. \label{fig.correlations}}
\end{figure}

The compressed method of quantum computation considered here can also be used to measure correlations (see also previous section). We focus here on those correlations which can be measured between two arbitrary qubits, say the $j$\th and the $k$\th. This correlation is quantified by the expectation value of the observable
\beq
    \label{eq.Cjk}
    C_{j,k}\equiv X_j Z_{j+1}\dots Z_{k-1}X_k= -ix_{2j+1}x_{2k}.
\eeq
As in the previous subsections, the observable which needs to be considered in the compressed circuit can easily be computed. It is given by
\beq
    \label{eq.barCjk}
    \overline{C}_{j,k}=-i\big(\kb{2j+1}{2k}-\kb{2k}{2j+1}\big).
\eeq
Therefore, the correlations between the $j$\th and the $k$\th qubits of a thermal state is given by
\beq
    \label{eq.Cjkcompressed}
    \expect{C_{j,k}}(\varphi,T)=n\tr\big[R(\varphi)\bar{\rho}(\varphi,T)[a]R^T(\varphi)\overline{C}_{j,k}\big],
\eeq
where $\bar{\rho}(\varphi,T)[a]$, is given in \eq{barrho} and $R(\varphi)$ is given in \eq{Rcircuit}.
In \figu{correlations}, we depict the correlations occurring in a system of $64$ qubits. In particular, the correlations between the central qubits, i.e. at position $\frac{n}{2}$ and the qubit at position $j$ is depicted as a function of $j$. To compute the curves depicted in \figu{correlations} (a), we have considered a fixed value $g=0.8$ and various temperatures between $T=0$ and $T=0.9$. The correlations decrease rapidly with distance and with temperature. Additionally, the correlations also decrease as the parameter $g$ increases. This is shown in \figu{correlations} (b), where each curve corresponds to a different value of $g$ for $T=0$.

\subsection{Construction of the compressed circuit for excited states}\label{sect.compressedexcited}

In this subsection we extend the compressed circuit presented above to the case where the measurements are performed on arbitrary eigenstates of $H(\varphi)[c]$, $\ket{\Psi_k(\varphi)}$ [see \eq{psik}]. In order to do so, let us define the operator
\beq
    \label{eq.Wk}
    W_k\equiv\prod_{j=0}^{n-1}X_j^{k_{n-1-j}},
\eeq
such that $\ket{k}=W_k\ket{0}^{\otimes{n}}$. Using \eq{psik} we can write the eigenstates of $H(\varphi)[c]$ as $\ket{\Psi_k(\varphi)}=U(\varphi)W_k\ket{0}^{\otimes n}$. The expectation value of the observable $A$ measured on the state $\ket{\Psi_k(\varphi)}$ can then be written as
\beq
    \expect{A}(\varphi,k)=\tr\left[U(\varphi)W_k\left(\proj{0}\right)^{\otimes n} W_k\adj{U}(\varphi)A\right].
\eeq

To obtain the compressed gate $R_{W_k}$ corresponding to the unitary $W_k$, we use a generalization of the procedure used to obtain the compressed gate $R_{B_0}$ in \eq{RB0}. That is, we write the conjugation of the Majorana operator $x_k$ by the Pauli operator $X_j$ as
\beq
    X_j x_k X_j=\sum_{l=0}^{2n-1} \left[R_{X_j}\right]_{k,l} x_l.
\eeq
Here, the matrix $R_{X_j}$ is diagonal, with diagonal entries $ \left[R_{X_j}\right]_{l,l}=1$ for $l\leq 2j$ and $ \left[R_{X_j}\right]_{l,l}=-1$ otherwise. Thus, the compressed gate $R_{W_k}=\prod_{j=0}^{n-1}R_{X_j}^{k_{n-1-j}}$ is also diagonal, with components
\beq
    \left[R_{W_k}\right]_{r,r}=
    \begin{cases}
        1& \text{for $r=1$,} \\
        (-1)^{\sum_{s=1}^{l}k_{n-s}} & \text{for $r=2l$ or $r=2l+1$.}\\
    \end{cases}
\eeq

The initial state of the compressed circuit can be computed in a similar way as $\bar{\rho}(g)[a]$ in \eq{barrho}. We find  $\bar{\rho}[a]=\frac{1}{2n}\left[\one-iS(\varphi)\right]$ [cf. \sect{compressedGibbs}], with $\left[S(\varphi)\right]_{r,s}\equiv\tr(-ix_r x_s \proj{0}^{\otimes n})$. This state can be written as \footnote
{
Note that the state $(\proj{0})^{\otimes n}$ is equal to the thermal state of $H(g\rightarrow\infty)[a]$ at $T=0$., so one could compute $\bar{\rho}[a]$ simply by taking the limit $\bar{\rho}(\varphi)[a]=\lim_{T\rightarrow 0}\bar{\rho}(\varphi,T)[a]$, with $\bar{\rho}(\varphi,T)[a]$ given by \eq{barrho}
}
\beq
    \label{eq.barrhoa}
    \bar{\rho}[a]=\frac{\one}{n}\otimes\ket{+_y}_m\bra{+_y}_m.
\eeq
Therefore, the expectation value of an observable $A$ measured on the excited state $\ket{\Psi_k(\varphi)}$, can be written as
\beq
    \label{eq.expectAgk}
    \expect{A}(\varphi,k)=n\tr\left[R(\varphi)R_{W_k}\bar{\rho}[a]R^T_{W_k}R^T(\varphi)\bar{A}\right].
\eeq

The magnetization, correlations $\nu$, or correlations $\expect{C_{i,j}}$ of any excited state $\ket{\Psi_k(g)}$ can be measured by choosing as observables in \eq{expectAgk} the operators $\overline{M}$, $\overline{K}$ and $\overline{C}_{i,j}$ respectively. For example, the magnetization $\expect{M}(\varphi,k)$ of an arbitrary excited state $\ket{\Psi_k(\varphi)}$ is given by
\beq
    \label{eq.expectMgk}
    \expect{M}(\varphi,k)=\tr\left[R(\varphi)R_{W_k}\bar{\rho}[a]R^T_{W_k}R^T(\varphi)Y_m\right].
\eeq
As before, the compressed circuit can be directly read off this expression.

\begin{figure}[ht]
\centering
	\includegraphics[width=0.4\textwidth]{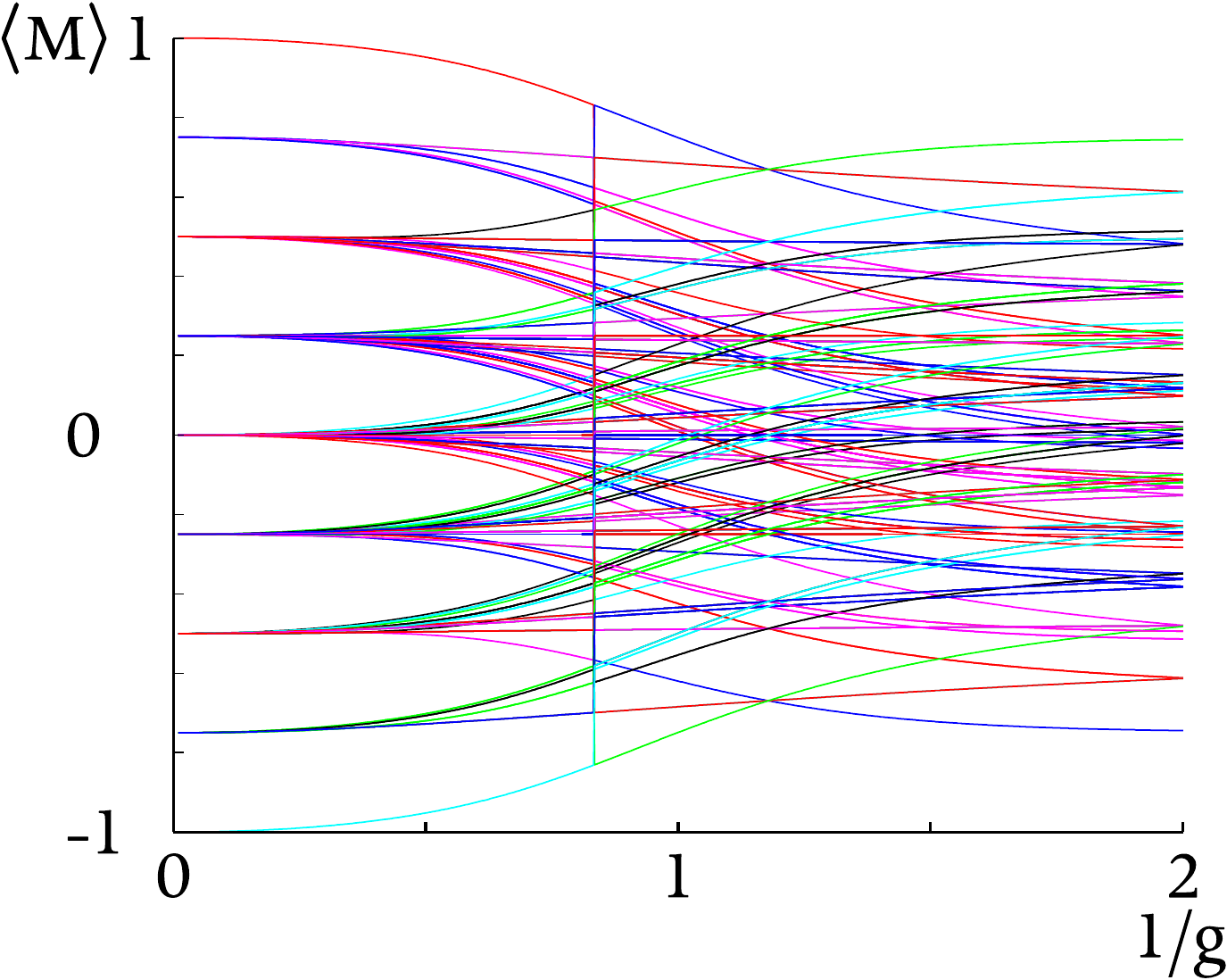}
    \caption{(Color online.) The magnetization for different eigenstates of the XY Hamiltonian as a function of $1/g$ and for $\delta=0.2$ and $n=8$ is shown. Due to the boundary conditions, several magnetization curves exhibit a discontinuity at the critical point (see \sect{preliminaries}), as indicated by a vertical line in the figure at the value $g_c^{-1}=(1+\delta)^{-1}=0.83$.\label{fig.magnetizationexcited}}
\end{figure}

In \figu{magnetizationexcited} the magnetization of each eigenstate of the XY Hamiltonian is shown. Each curve corresponding to a given eigenstate is obtained by computing $\expect{M}(\varphi,k)$ as a function of $1/g$ for a fixed value of $k$ and $\delta$. Given that the number of curves grows exponentially with the system size $n$, we have considered here a small system size of $n=8$ qubits. Let us note that due to the level crossing at the critical point $g_c$, the magnetizations have a discontinuity at this point, and the curves corresponding to different excited states cross each other (cf. \figu{magnetization}).

\subsection{Time evolution}

The product of matchgates $U$ recalled in \sect{decomposition} can also be used to decompose the time-evolution operator $V(t)=\epower^{-iHt}$ into a product of matchgates. In this way it is possible to simulate the measurement outcome of a time-evolved initial state with a matchgate circuit. In the following we briefly discuss the construction of this matchgate circuit and of the equivalent compressed circuit.

Let us first consider the initial state $\left(\proj{0}\right)^{\otimes n}$; the output of the matchgate circuit described above is thus given by
\beq
    \label{eq.expectAtimeevolution}
    \expect{A}(t)=\tr\left[V(t)\left(\proj{0}\right)^{\otimes n}\adj{V}(t) A\right].
\eeq
As in the previous subsections, we assume that the observable $A$ is of the form given by \eq{Aquadratic}. Due to \eq{UHU} we have $V(t)=UW(t)\adj{U}$, where $W(t)=\epower^{-iH[a]t}$. The diagonal Hamiltonian $H[a]$ as a function of Pauli operators is given in \eq{Haexpansion}. It follows that $W(t)$ can be decomposed as the following product of matchgates
\beq
    \label{eq.Wt}
    W(t)=\prod_{j=0}^{2n-1}\epower^{-i\frac{\epsilon_j}{2} Z_j t}.
\eeq

In order to obtain the compressed gate associated to $W(t)$, we write $H[a]$ as a quadratic polynomial in the Majorana operators using the Jordan-Wigner representation given in \eq{JordanWigner} and obtain $H[a]=i\sum_{j=0}^{n-1}\frac{\epsilon_j}{4}\left(x_{2j}x_{2j+1}-x_{2j+1}x_{2j}\right)$. From this expression, and following the procedure described in \sect{preliminaries} one can see that the compressed matrix $R_W(t)$ corresponding to $W(t)$ is given by
\beq
    \label{eq.RWt}
    R_W(t)=\bigoplus_{j=0}^{n-1}
        \bpmat
            \phantom{-}\cos(\epsilon_j t) & \sin(\epsilon_j t)\\
            -\sin(\epsilon_j t) & \cos(\epsilon_j t)\\
        \epmat.
\eeq

In the same way as in the previous subsections, we rewrite the expectation value $\expect{A}(t)$ as
\beq
    \label{eq.expectAt}
    \expect{A}(t)=\tr\big([RR_W(t)R^T\bar{\rho}_{\mathrm{in}}RR^T_W(t)R^T\bar{A}\big),
\eeq
where $\bar{\rho}_{\mathrm{in}}=\frac{\one}{n}\otimes\ket{+_y}_m\bra{+_y}_m$, and the observable $\bar{A}$ can be computed from $A$ using \eq{overlineA}. Thus, we have that a compressed quantum circuit comprising of the preparation of the initial state $\bar{\rho}_{\mathrm{in}}$, the evolution of the system under the action of the orthogonal matrix $RR_W(t)R^T$ and the measurement of the observable $\bar{A}$ can be used to compute the expectation value $\expect{A}(t)$  given in \eq{expectAtimeevolution}.

The matchgate circuit considered above can be easily generalized to include any initial state that can be efficiently generated with matchgates from the state $\left(\proj{0}\right)^{\otimes n}$. To obtain such a circuit it is sufficient to insert at the beginning of the matchgate circuit described above the matchgates that prepare the desired initial state. The compressed circuit has to be modified accordingly by the set of compressed gates corresponding to the additional matchgates.

A more general set of initial states can be considered in the case where the restriction to a logarithmical space classical computer is removed. In this case, one can consider an arbitrary initial state of the form
$\rhoin=\rho_0\otimes\dots\otimes \rho_{n-1}$, where each $\rho_i$ is a density matrix of a single qubit. The initial state that one has to consider in the equivalent compressed circuit is given by $\bar{\rho}_{\mathrm{in}}\propto\left[\alpha\one-iS(\rho)\right]$, where $S(\rho)$ is defined by $\left[S(\rho)\right]_{j,k}\equiv\tr\left(-i x_j x_k \rho\right)$, where the Majorana operators $x_j$ are given in \eq{JordanWigner}. The real parameter $\alpha$ has to be chosen in such a way that $\bar{\rho}_{\mathrm{in}}$ is positive-semidefinite. Note that the preparation of $\bar{\rho}_{\mathrm{in}}$ requires the classical storage of all the components of $S(\rho)$, which is of dimension $n^2$.

Let us note that in \cite{BoMu12} we presented a compressed quantum circuit that can be used to simulate the time evolution of a XY chain of qubits. In contrast to the circuit presented here, in \cite{BoMu12} the compressed quantum circuit was constructed starting from a Trotter decomposition of the time-evolution operator leading to certain errors in the estimation of $\expect{A}(t)$ that depend on the simulated time $t$. Here, by using the matrix $U$ that exactly diagonalizes the XY Hamiltonian, $\expect{A}(t)$ can be computed exactly for any value of $t$.

\section{Conclusion}\label{sect.conclusion}

We derived compressed quantum circuits to measure physically interesting properties of the 1D--XY--model. In order to do so, we used the decomposition into a product of matchgates of the unitary $U$ that diagonalizes the XY Hamiltonian, and the fact that any eigenstate or thermal state of the system can be prepared by evolving a product state according to $U$ \cite{LaVe08}. Due to the equivalence between matchgate circuits and compressed quantum computation \cite{JoKr10}, the corresponding matchgate circuits can be compressed as long as the measurement operator is quadratic in the Majorana operators. That is, there is a compressed quantum circuit that simulates the matchgate circuit, running on exponentially less qubits. In the cases considered here, we have additionally shown that the compressed circuit requires not only less qubits, but also less gates than the matchgate circuit that it simulates. Let us finally mention that the results presented here can be extended to any quadratic Hamiltonian. The reason for that is that the matrix that diagonalizes any of these Hamiltonians can always be decomposed into a polynomial number of matchgates \cite{jm08}.

\section{Acknowledgments}\label{sect.acknowledgments}

This research was funded by the Austrian Science Fund (FWF), grant no. Y535-N16.

\begin{appendices}

\section{Reordering transformations}\label{ap.appendixA}

The aim of this appendix is to provide a more detailed description of the product of matchgates corresponding to the various reordering operations mentioned in \sect{reordering} [see \eq{mathcalU}]. Let us recall that each of these transformations can be decomposed into elementary transformations $\mathcal{S}_{j,k}$, which map the pair of fermionic operators $(a_j,a_k)$ into the pair $(a_k,a_j)$. The transformations $\mathcal{S}_{j,k}$ has the associated matchgate $S_{j,k}$ given in \eq{Sjk}. Therefore, the product of matchgates corresponding to an arbitrary reordering transformation $\mathcal{S}$ can be easily obtained from the decomposition of $\mathcal{S}$ in terms of transformations $\mathcal{S}_{j,k}$.

A convenient way of providing such a decomposition is by associating to the transformation $\mathcal{S}$ an operation that permutes the components of the vector $\boldsymbol{\mu}$ introduced in \eq{psi2} in \sect{decomposition}. For example, let us consider the transformation $\mathcal{S}_{j,k}$  that maps the pair of operators $(a_j,a_k)$ into the pair $(a_k,a_j)$. This transformation has an associated unitary matrix $S_{j,k}$ that transforms an arbitrary state $\ket{\Psi[a,\boldsymbol{\mu}]}$ into a new state $S_{j,k}\ket{\Psi[a,\boldsymbol{\mu}]}=\ket{\Psi[a,\boldsymbol{\mu'}]}$ [see \eq{psi2}], where $\mu'_j=\mu_k$, $\mu'_k=\mu_j$ and $\mu'_l=\mu_l$, $\forall l\neq j,k$. Therefore, we can associate to the transformation $\mathcal{S}_{j,k}$ the operation that permutes the components $j$ and $k$ of the vector $\boldsymbol{\mu}$. In the subsequent sections we apply this procedure to decompose the transformations  $\mathcal{S}^{(s)}$, for $0\leq s\leq m$ and $\mathcal{S}_{\mathrm{Bog}}$ in terms of elementary transformations of the form $\mathcal{S}_{j,k}$.

\subsection*{Reordering transformation $\mathcal{S}^{\mathrm{Bog}}$}

The reordering transformation $\mathcal{S}^{\mathrm{Bog}}$ is the first step of the transformation $\mathcal{U}$ given in \eq{mathcalU}. Therefore, the unitary matrix $S^{\mathrm{Bog}}$  associated with this transformation acts directly on the initial state of the quantum circuit. The quantum initial state can be associated to fermionic operators $a_j$ according to \eq{psi2}. This association establishes a particular ordering of the operators $a_j$ that is described by a vector $\boldsymbol{\lambda}^{(0)}=\left(0,\dots,n-1\right)$.

The reordering transformation $\mathcal{S}^{\mathrm{Bog}}$ precedes the Bogoliuvov transformation $\mathcal{B}$, given in \sect{bogoliuvovtransformation}. The latter maps the operators $\{a_i\}_{i=0,\dots,n-1}$ into the operators $\{b_i\}_{i=0,\dots,n-1}$, by mixing the pair of operators $a_j$ with $a_{-j}$, for $-j\equiv n-j$, and $a_0$ with $a_{\frac{n}{2}}$. In order that these operators act on consecutive qubits [according to \eq{psi2}], the matrix $S^{\mathrm{Bog}}$ has to transform the vector $\boldsymbol{\lambda}^{(0)}$ into some vector $\boldsymbol{\lambda}^{\mathrm{Bog}}$ whose components are paired in the form $(j,-j)$ and $(0,\frac{n}{2})$. This condition still allows for multiple choices of the vector $\boldsymbol{\lambda}^{\mathrm{Bog}}$. In this work we have chosen a particular ordering with the aim of obtaining a simple expression for the compressed circuit associated to the matrix $S^{\mathrm{Bog}}$. This ordering is given by the vector
\begin{multline}
    \boldsymbol{\lambda}^{\mathrm{Bog}}=\Big(0,\frac{n}{2},2,-2,\dots 2l,-2l,\dots,\frac{n}{2}-2,-\frac{n}{2}+2,\\
    \frac{n}{2}-1,-\frac{n}{2}+1,\dots, 2l-1,-2l+1,\dots,1,-1\Big).
\end{multline}

It is simple to verify that $\boldsymbol{\lambda}^{\mathrm{Bog}}$ can be obtained from $\boldsymbol{\lambda}^{(0)}$ by permuting successively the components $\sigma_0\equiv\frac{n}{2}$ and $\sigma_j$, where
\beq
    \label{eq.sigma_j}
    \sigma_j=\frac{n}{2}+(-1)^j j ,\quad \text{for $j=0,\dots,\frac{n}{2}-1$}.
\eeq
Given that the permutation of two components, e.g. $\sigma_0$ and $\sigma_j$ is associated with the fermionic swap gate $S_{\sigma_0,\sigma_j}$, it follows that the unitary matrix corresponding to the transformation $\mathcal{S}^{\mathrm{Bog}}$ is given by the following product of matchgates
\beq
    \label{eq.SBog}
    S^{\mathrm{Bog}}=\prod_{j=1}^{\frac{n}{2}-1} S_{\sigma_0,\sigma_j}.
\eeq

\subsection*{Reordering transformation $\mathcal{S}^{(s)}$ for $1\leq s \leq m-1$}

The reordering transformation $\mathcal{S}^{(s)}$ acts after the transformation $\mathcal{F}^{(s-1)}$ and before the transformation $\mathcal{F}^{(s)}$ presented in \sect{fouriertransformation}. Recall that the transformation $\mathcal{F}^{(s)}$ maps a set of operators $\{x_i^{(s)}\}_{i=0,\dots,n-1}$ into the set $\{x_i^{(s+1)}\}_{i=0,\dots,n-1}$, by mixing pairs of operators whose indices differ solely on the  $(m-s-1)$\th component. The matrix $F^{(s)}$ associated with the transformation $\mathcal{F}^{(s)}$ acts on neighboring qubits in case where the association between qubits and fermionic operators is given by the vector $\boldsymbol{\lambda}^{(s)}$ introduced in \sect{fouriertransformation}. Let us recall that the components of this vector are given by $\lambda^{(s)}_{2l}=r_{l,s}$ and $\lambda^{(s)}_{2l+1}=r_{l,s}+2^{m-s-1}$, where $r_{l,s}$ is given in \eq{rl}. Therefore, the reordering transformation $\mathcal{S}^{(s)}$ has to be associated with the operation that maps the vector $\boldsymbol{\lambda}^{(s-1)}$ into the vector $\boldsymbol{\lambda}^{(s)}$.

In order to see which components of $\boldsymbol{\lambda}^{(s-1)}$ have to be permuted to obtain the vector  $\boldsymbol{\lambda}^{(s)}$, we now write down the binary decomposition of the elements of these two vectors. Using \eq{rl} we have that
\beq
    \label{eq.listoflambdas}
    \bal
        \left[\lambda^{(s-1)}_{2l}\right]&= \begin{bmatrix} l_{m-2}\dots l_{m-s+1} & 0 & l_{m-s}& l_{m-s-1} \dots l_0\end{bmatrix},\\
        \left[\lambda^{(s-1)}_{2l+1}\right]&=\begin{bmatrix}l_{m-2}\dots l_{m-s+1}& 1 & l_{m-s}& l_{m-s-1} \dots l_0\end{bmatrix},\\
        \left[\lambda^{(s)}_{2l}\right]&=\begin{bmatrix}l_{m-2}\dots l_{m-s+1}& l_{m-s} & 0 & l_{m-s-1} \dots l_0\end{bmatrix},\\
        \left[\lambda^{(s)}_{2l+1}\right]&=\begin{bmatrix}l_{m-2}\dots l_{m-s+1} & l_{m-s}& 1 &  l_{m-s-1} \dots l_0\end{bmatrix}.
    \eal
\eeq
One can see that only half of the components have to be reordered and therefore, $\frac{n}{4}$ permutations are required. Let us consider, for example, the $2l$\th component. For a value of $l$ where $l_{m-s}=0$, we have that $\lambda_{2l}^{(s-1)}=\lambda_{2l}^{(s)}$, thus this component does not have to be reordered. The same happens with the $(2l+1)$\th component when $l_{m-s}=1$. The only components that have to be permuted are the $(2l)$\th and the $(2l'+1)$\th, in case where $l$ is such that $l_{m-s}=1$ and $l'=l\oplus 2^{m-2}$. Note that this is where the binary decomposition of $l'$ is equal to the one of $l$ for any bit except the $(m-s)$\th. This can be written in a compact way as follows. Two components $z_1(r)^{(s)}$ and $z_2(r)^{(s)}$ of the vector $\boldsymbol{\lambda}^{(s-1)}$ must to be permuted if they take the form
\beq
    \label{eq.jk}
    \bal
        \left[z_1(r)^{(s)}\right]&=&[r_{m-3},\dots, r_{m-s}, &1, r_{m-s-1}, \dots, r_0,0],\\
        \left[z_2(r)^{(s)}\right]&=&[r_{m-3},\dots, r_{m-s}, &0, r_{m-s-1}, \dots, r_0,1],\\
    \eal
\eeq
for some $0\leq r\leq \frac{n}{4}-1$ with a binary decomposition given by $[r]=[r_{m-3},\dots,r_0]$.

In the following we define the set
\beq
    \Omega_{S^{(s)}}\equiv\Big\{\Big(z_1(r)^{(s)},z_2(r)^{(s)}\Big)\Big\}_{r=0,\dots\frac{n}{4}-1},
\eeq
where  $z_1(r)^{(s)}$ and $z_2(r)^{(s)}$ are given in \eq{jk}. Note that the set $\Omega_{S^{(s)}}$ groups the pair of indices that have to be permuted with each other in order to transform the vector $\boldsymbol{\lambda}^{(s-1)}$ into $\boldsymbol{\lambda}^{(s)}$. Using this definition, it follows that the product of matchgates associated to the transformation  $\mathcal{S}^{(s)}$ is given by the product of fermionic swap gates
\beq
    \label{eq.Ss}
    S^{(s)}=\prod_{(j,k)\in \Omega_{S^{(s)}}}S_{j,k}.
\eeq

\subsection*{Reordering transformation $\mathcal{S}^{(0)}$}

The reordering transformation $\mathcal{S}^{(0)}$ acts after the transformation $\mathcal{S}_{\mathrm{Bog}}^{-1}$  and before the transformation $\mathcal{F}^{(0)}$ [see \eq{mathcalU}]. In the same way as in the previous subsections we associate with the transformation $\mathcal{S}^{(0)}$ an operation that permutes the components of the vector $\boldsymbol{\mu}$, whose components indicate which qubit is associated with  which fermionic operator.

Given that the reordering transformation $\mathcal{S}_{\mathrm{Bog}}^{-1}$ inverts the reordering  produced by $\mathcal{S}_{\mathrm{Bog}}$, after this transformation, the association between qubits and fermionic operators is given by the vector $\boldsymbol{\lambda}^{(0)}$. However, in order for the transformation $\mathcal{F}^{(0)}$  to act on pairs of operators with consecutive indices, it is necessary that the association between qubits and fermionic operators is given by a vector $\boldsymbol{\lambda}^{(1)}$ [see \sect{fouriertransformation}]. Therefore, the transformation $\mathcal{S}^{(0)}$ must correspond to the transformation of the vector $\boldsymbol{\lambda}^{(0)}$ into the vector $\boldsymbol{\lambda}^{(1)}$. Using \eq{rl} we can write the binary decomposition of the components of these two vectors as
\beq
    \label{eq.listoflambdas2}
    \bal
        \left[\lambda^{(0)}_{2l}\right]&=[l_{m-2}\dots l_0\;0]\\
        \left[\lambda^{(0)}_{2l+1}\right]&=[l_{m-2}\dots l_0\;1]\\
        \left[\lambda^{(1)}_{2l}\right]&=[0\;l_{m-2}\dots l_0]\\
        \left[\lambda^{(1)}_{2l+1}\right]&=[1\;l_{m-2}\dots l_0].\\
    \eal
\eeq
In order to obtain the permutations required to transform $\boldsymbol{\lambda}^{(0)}$ into $\boldsymbol{\lambda}^{(1)}$, we can use the results of the previous section. Comparing the \eq{listoflambdas2} to  \eq{listoflambdas}, we see that the transformation $\mathcal{S}^{(0)}$ can be obtained as a concatenation of the transformations $\mathcal{S}^{(s)}$, with $s$ from $m-1$ to $1$. It follows that the unitary matrix associated with the transformation $\mathcal{S}^{(0)}$  is given by the product
\beq
    \label{eq.S0}
    S^{(0)}=S^{(1)}\dots S^{(m-1)},
\eeq
where the gates $S^{(s)}$ for $1\leq s\leq m-1$ are given in \eq{Ss}.

\subsection*{Reordering transformation $\mathcal{S}^{(m)}$}

The reordering transformation $\mathcal{S}^{(m)}$ acts after the transformation $\mathcal{F}^{(m-1)}$ [see \sect{fouriertransformation}]. Therefore, the unitary matrix $S^{(m)}$ corresponding to this transformation yields the output state of the circuit. In the following we see how to obtain the permuting operation of the vector $\boldsymbol{\mu}$ associated with $S^{(m)}$.

Let us stress that in both the input and the output states, the association between qubits and fermionic operators $a_j$ and $c_j$ respectively, must be given by the same vector $\boldsymbol{\lambda}^{(0)}$. Note however that the matrix $F^{(m-1)}$ yields states where the association between qubits and the fermionic operators $x^{(m-1)}_j$ is given by the vector $\boldsymbol{\lambda}^{(m-1)}$ [see \sect{fouriertransformation}]. That means, using the notation introduced in \eq{psi2}, that the output state is a state of the form $\ket{\psi[x^{(m-1)},\lambda^{(m)}]}$. However, the operators $x_j^{(m-1)}$ are related to the operators $c_j$ according to \eq{ckequivalence}, i.e., $x^{(m-1)}_j=c_{j'}$, with $[j]=[j_{m-1},\dots,j_0]$ and $[j']=[j_{0},\dots,j_{m-1}]$. Therefore, a state $\ket{\Psi[x^{(m-1)},\boldsymbol{\lambda}^{(m)}]}$ is equal to the state $\ket{\Psi[c,\boldsymbol{\lambda}^{'}]}$, where the binary decomposition of the components of the vector $\boldsymbol{\lambda}^{'}$ is given by
\beq
    \label{eq.listoflambdas3}
    \left[\lambda^{'}_{j}\right]=[j_0,\dots ,j_{m-1}].
\eeq

From the discussion above, it follows that the matrix $S^{(m)}$ must be associated with the operation that transforms the vector $\boldsymbol{\lambda}^{'}$ into the vector $\boldsymbol{\lambda}^{(0)}$ with $\left[\lambda^{(0)}_j\right]=\left[j_{m-1},\dots,j_0\right]$. This transformation can be achieved as follows. Let us consider a transformation that maps the vector $\boldsymbol{\lambda}'$ to another vector $\tilde{\boldsymbol{\lambda}}'$ in such a way that the $t$\th and $(m-t-1)$\th binary components of the elements of $\left[\tilde{\lambda}'_j\right]$ are permuted. That is $\left[\tilde{\lambda}'_{j}\right]=\left[j_0,\dots, j_{t-1},j_{m-t-1},j_{t+1},\dots,j_{m-t-2},j_{t},j_{m-t},\dots,j_{m-1}\right]$. Applying these transformation for $0\leq t\leq t_{\mathrm{max}}$, with $t_{\mathrm{max}} = \lfloor m/2 \rfloor$ to the vector $\boldsymbol{\lambda}'$ yields the vector $\boldsymbol{\lambda}^{(0)}$. Denoting by $T^{(t)}$ for $0\leq t\leq t_{\mathrm{max}}$ the product of matchgates associated with each of the aforementioned transformations, it follows that the matrix $S^{(m)}$ can be written as
\beq
    \label{eq.Sm}
    S^{(m)}=T^{(t_{\mathrm{max}})} \cdots T^{(0)}.
\eeq

We now describe how to obtain the decomposition into matchgates of each of the unitaries $T^{(t)}$. From the discussion above we have that $T^{(t)}$ is associated with a transformation of the vector $\boldsymbol{\lambda}'$ that permutes two components $y_1^{(t)}(r)$ and $y_2^{(t)}(r)$ if and only if they are of the form
\beq
    \label{eq.jk2}
    \bal
        \left[y_1^{(t)}(r)\right]=&[r_{m-3},\dots, r_{m-4-t},0,r_{m-3-t},\dots,r_{t},1,\\
        &\phantom{[} r_{t-1},\dots,r_0],\\
        \left[y_2^{(t)}(r)\right]=&[r_{m-3},\dots, r_{m-4-t},1,r_{m-3-t},\dots,r_{t},0,\\
        &\phantom{[} r_{t-1},\dots,r_0],\\
    \eal
\eeq
for any $0\leq r \leq \frac{n}{4}$-1. Defining the set
\beq
    \Omega_{T^{(t)}}\equiv\Big\{\big[y_1^{(t)}(r),y_2^{(t)}(r)\big]\Big\}_{r=0,\dots \frac{n}{4}-1},
\eeq
where $y_1^{(t)}(r)$ and $y_2^{(t)}(r)$ are given in \eq{jk2}, it follows that the matchgate $T^{(t)}$ can be written as
\beq
    \label{eq.Tr}
    T^{(t)}=\prod_{(j,k)\in \Omega_{T^{(t)}}}S_{j,k}.
\eeq

With this transformation, we conclude the construction of the product of matchgates corresponding to each of the reordering transformations. In the following appendix, we provide the compressed orthogonal gates corresponding to each of these product of matchgates.

\section{Compression of the reordering transformations}\label{ap.appendixB}

In Appendix \ref{ap.appendixA} we presented unitary matrices acting on quantum states that are associated to the reordering transformations used in the diagonalization of the Hamiltonian. Moreover, we provided a decomposition of these matrices as a product of matchgates $S_{j,k}$. In this section, we provide the compressed orthogonal matrices that correspond to these products of matchgates. The compressed gate corresponding to $S_{j,k}$, is given by $R_{S_{j,k}}$ [see \eq{RSjk}]. Therefore, a direct way to obtain the compressed gates corresponding to the product of matchgates $S^{\mathrm{Bog}}$, $S^{(0)}$, $S^{(s)}$ for $1\leq s\leq m-1$ and $S^{(m)}$ is to use the same decompositions given in \eq{SBog}, \eq{S0}, \eq{Ss} and \eq{Sm} respectively, in terms of the matrices $R_{S_{j,k}}$, instead of the matchgates $S_{j,k}$. One can easily verify that these decompositions require $\Order\left[\mathrm{poly}(n)\right]$ matchgates. However, due to the symmetries of the problem, it is possible to rewrite the compressed gates corresponding to entire reordering steps with at most $\Order\left[\log(n)^2\right]$ elementary gates, as we show within this appendix.

\subsection*{Compression of the reordering transformation $\mathcal{S}^{(s)}$, for $1\leq s\leq m-1$}

Here, we consider the construction of the compressed matrix $R_{S^{(s)}}$, corresponding to the product of matchgates $S^{(s)}$, that has been given as a decomposition in term of fermionic swap gates in \eq{Ss}, for $ 1\leq s\leq m-1$. As we pointed out above, we can decompose $R_{S^{(s)}}$ in term of matrices $R_{S_{j,k}}$ using the decomposition given in \eq{Ss}, that is
\beq
    \label{eq.RSs1}
    R_{S^{(s)}}=\prod_{(j,k)\in \Omega_{S^{(s)}}} R_{S_{j,k}}.
\eeq
Using the explicit form of the matrix $R_{S_{j,k}}$ given in \eq{RSjk}, and the fact that none of the indices $j$ occurs more than once as an element of a pair in $\Omega_{S^{(s)}}$, \eq{RSs1} can be written explicitly as $R_{S^{(s)}}=\left[\one+\sum_{(j,k)\in \Omega^{(s)}}\big(\kb{j}{k}+\kb{k}{j}-\proj{j}-\proj{k}\big)\right]\otimes\one_m$. This expression can be written in a more compact way by using the binary decomposition of the indices $(j,k)$ in the set $\Omega_{S^{(s)}}$, i.e., using \eq{jk} leads to
\beq
    \label{eq.RSs}
    \bal
         R_{S^{(s)}}&=\Big[\one+\sum_{r=0}^{\frac{n}{4}-1}\proj{r}\otimes\big(\kb{01}{10}+\kb{10}{01}\\
         &\phantom{=}-\proj{01}-\proj{10}\big)_{s-1,m-1}\Big]\otimes\one_m\\
         &=\big(\proj{00}+\proj{11}\\
         &\phantom{=}+\kb{10}{01}+\kb{01}{10}\big)_{s-1,m-1}\otimes\one,
    \eal
\eeq
where the identity operator acts on all qubits except on the $(s-1)$\th and the $(m-1)$\th. Note that $R_{S^{(s)}}$ is the swap gate between the qubits $s-1$ and $m-1$, i.e. a single two-qubit gate.

\subsection*{Compression of the reordering transformation $\mathcal{S}^{(0)}$}

Due to \eq{S0} we have that the compressed gate $R_{S^{(0)}}$ can be written as
\beq
    \label{eq.RS0}
    R_{S^{(0)}}=R_{S^{(1)}}\cdots R_{S^{(m-1)}},
\eeq
where the matrices $R_{S^{(s)}}$, for $1\leq s\leq m-1$ are two-qubit gates, given in \eq{RSs}. Therefore, $R_{S^{(0)}}$ is a product of $m-1$ elementary gates.

\subsection*{Compression of the reordering transformation $\mathcal{S}^{(m)}$}

To construct $R_{S^{(m)}}$ we need the compressed gates $R_{T^{(t)}}$, corresponding to the product of matchgates given in \eq{Tr}. Given the similarity between the definitions of $T^{(t)}$ and $S^{(s)}$, the matrix $R_{T^{(t)}}$ can be constructed analogously to $R_{S^{(s)}}$. We obtain
\beq
    \bal
        R_{T^{(t)}}=&\big(\proj{00}+\proj{11}\\
        &+\kb{10}{01}+\kb{01}{10}\big)_{t,m-1-t}\otimes\one.
    \eal
\eeq
Hence, $R_{T^{(t)}}$ is the swap gate between the qubits $t$ and $m-1-t$. Due to \eq{Sm} it follows that
\beq
    \label{eq.RSm}
    R_{S^{(m)}}=R_{T^{(t_{\mathrm{max}})}}\cdots R_{T^{(0)}}.
\eeq
As each matrix $R_{T^{(t)}}$ is a two-qubit gate, the implementation of $R_{S^{(m)}}$ requires $t_{\mathrm{max}}=\lfloor m/2 \rfloor$ two-qubit gates.

\subsection*{Compression of the reordering transformation $\mathcal{S}^{\mathrm{Bog}}$}

From the decomposition of $S^{\mathrm{Bog}}$ given in \eq{SBog}, it follows that its corresponding compressed matrix can be written as $R_{S^{\mathrm{Bog}}}=\prod_{j=1}^{\frac{n}{2}-1} R_{S_{\sigma_0,\sigma_j}}$, where the indices $\sigma_j$ are defined in \eq{sigma_j}. Using the explicit expression of the indices $\sigma_j$ and the matrices $R_{S_{j,k}}$ given in \eq{RSjk}, it is possible to express $R_{S^{\mathrm{Bog}}}$ in a more compact way. To do so, note that the product of two matrices $R_{\sigma_0,\sigma_j}$ and $R_{\sigma_0,\sigma_k}$ takes the form
\begin{multline}
    R_{S_{\sigma_0,\sigma_j}}R_{S_{\sigma_0,\sigma_k}}=\big(\one-\kb{\sigma_0}{\sigma_0}-\kb{\sigma_j}{\sigma_j}-\kb{\sigma_k}{\sigma_k}\\
    +\kb{\sigma_0}{\sigma_j}+\kb{\sigma_j}{\sigma_k}+\kb{\sigma_k}{\sigma_0}\big)\otimes\one_m.
\end{multline}
Generalizing this equation to the case where we have $\frac{n}{2}-1$ factors we obtain
\beq
    \label{eq.RSBog}
    R_{S^{\mathrm{Bog}}}=\left(\sum_{l\notin \Sigma}\kb{l}{l}+\sum_{i=0}^{\frac{n}{2}-1}\kb{\sigma_i}{\sigma_{i+1}}+\kb{\sigma_{\frac{n}{2}-1}}{\sigma_0}\right)\otimes\one_m,
\eeq
where we have defined $\Sigma=\{\sigma_j\}_{j=0,\dots n-1}$. Using the binary decomposition of the indices $\sigma_j$, it is possible to rewrite the previous expression in the following more compact way
\begin{multline}
    \label{eq.RSBog}
    R_{S^{\mathrm{Bog}}}=\Big[\big(\kb{00}{00}_{0,m-1}+\kb{11}{11}_{0,m-1}\big))\otimes\one\\
    +\kb{10}{01}_{0,m-1}\otimes A+\kb{01}{10}_{0,m-1}\otimes BA\Big]\otimes\one_m,
\end{multline}
where the operators $A$ and $B$ act on qubits $1,\dots, m-2$ and are given by
\beq
    \bal
        A&=\sum_{k=0}^{k_{\mathrm{max}}} \kb{k}{k_{\mathrm{max}}-k},\\
        B&=\sum_{k=0}^{k_{\mathrm{max}}-1} \kb{k+1}{k}+\kb{0}{k_{\mathrm{max}}},
    \eal
\eeq
with $k_{\mathrm{max}}=n/4-1$. Note that $A=X^{\otimes m-2}$, is a product of $m-1$ single-qubit gates. The operator $B$ acts on the computational basis state as $B\ket{j}=\ket{j\oplus1}$ where $\oplus$ is the addition modulo $2^{m-2}$ and can be decomposed as a product of ${m-2}$ controlled operations. More explicitly, we have that $B=X_{m-2}\left[\Lambda^{(m-2)}\left(X_{m-3}\right)\right]\dots\left[\Lambda^{(m-2),\dots,2}\left(X_{1}\right)\right]$, where $\Lambda^{i_1\dots,i_l}\left(O_k\right)$ denotes the single-qubit controlled operation $O$ acting on the qubit $k$, and the $i_1,\dots i_n$ are the control qubits. Given that any controlled operation acting on $m$ qubits can be decomposed into a product of $\Order(m)$ elementary gates \cite{BaBe95}, it follows that $R_{S^{\mathrm{Bog}}}$ given in \eq{RSBog} can be decomposed into a product of $\Order(m^2)$ elementary gates.

\end{appendices}

\bibliographystyle{ieeetr}

\end{document}